\documentclass[reprint,twocolumn,superscriptaddress,secnumarabic,amssymb, nobibnotes, aps,pra]{revtex4-1}

\usepackage{amsmath}    
\usepackage{graphicx}    
\usepackage{verbatim}    
\usepackage{color}           
\usepackage{subfigure}   
\usepackage{hyperref}    
\usepackage{verbatim}  
\usepackage[normalem]{ulem}

\begin{document}
\title{Core-Level X-Ray Spectroscopy of Infinite-Layer Nickelate:~LDA+DMFT Study}

\author{Keisuke Higashi}
\affiliation{Department of Physics and Electronics, Graduate School of Engineering, Osaka Prefecture University 1-1 Gakuen-cho, Nakaku, Sakai, Osaka 599-8531, Japan}
\author{Mathias Winder}
\affiliation{Institute for Solid State Physics, TU Wien, 1040 Vienna, Austria}
\author{Jan Kune\v{s}}
\affiliation{Institute for Solid State Physics, TU Wien, 1040 Vienna, Austria}
\author{Atsushi Hariki}
\affiliation{Department of Physics and Electronics, Graduate School of Engineering, Osaka Prefecture University 1-1 Gakuen-cho, Nakaku, Sakai, Osaka 599-8531, Japan}
\thanks{hariki@pe.osakafu-u.ac.jp}

\date{\today}

\date{\today}

\begin{abstract}
Motivated by recent core-level x-ray photoemission spectroscopy, x-ray absorption spectroscopy (XAS), and resonant inelastic x-ray scattering (RIXS) experiments for the newly discovered superconducting infinite-layer nickelate, we investigate the core-level spectra of the parent compounds NdNiO$_2$ and LaNiO$_2$ using the combination of local density approximation and dynamical mean-field theory (LDA+DMFT).
Adjusting a charge-transfer energy to match the experimental spectra, we determine the optimal model parameters and discuss the nature of the NdNiO$_2$ ground state. 
We find that self-doping from the Nd 5$d$ states in the vicinity of the Fermi energy prohibits opening of a Mott-Hubbard gap in NdNiO$_2$.
The present Ni $L_3$ XAS and RIXS calculation for LaNiO$_2$ cannot explain the difference from NdNiO$_2$ spectra.
\end{abstract}

\maketitle

\section{Introduction}

High-$T_c$ superconductivity of cuprates has been a focal point of 3$d$ transition-metal oxide (TMO) physics over the past 30 years~\cite{Bednorz86,imada98,Dagotto94}; nevertheless, the underlying mechanism remains elusive.
Superconductivity~\cite{li19} reported recently in layered nickelate Nd$_{0.8}$Sr$_{0.2}$NiO$_2$ ($T_c$ = 9--15~K) with a similar crystal structure
may provide new clues.
The fundamental question 
is whether
the electronic structure
of NdNiO$_2$ (and LaNiO$_2$) is similar to that of high-$T_c$ cuprates. 
Naively, one might presume that 
Ni in the undoped systems is monovalent and, thus, hosts the $d^9$ ($S=1/2$) ground state similar to cuprates.
However, theoretical studies~\cite{Botana20,Krishna20,Lee04,Zhang20,Hepting19} suggest a self-doping from Nd (or La) 5$d$ orbitals. 
Additionally, holes doped to a low-valence Ni$^{1+}$ compound may reside in Ni 3$d$ orbitals, unlike in cuprates~\cite{Zaanen85,imada98,Dagotto94} or NiO with Ni$^{2+}$~\cite{kunes07b},
where they occupy the O 2$p$ states. 


The Ni 2$p_{3/2}$ core-level x-ray photoemission spectroscopy (XPS)~\cite{Fu19}, x-ray absorption spectroscopy (XAS), and resonant inelastic x-ray scattering (RIXS)~\cite{Hepting19,Rossi20}
are employed to probe the electronic structure of infinite-layer nickelates.
A shoulder observed
in the main line of the Ni 2$p_{3/2}$ XPS spectra in NdNiO$_2$~\cite{Fu19}
is attributed to Ni-Ni charge-transfer (CT) response to the creation of the core hole, 
a process traditionally called nonlocal screening (NLS)~\cite{veenendaal93}.
Generally, NLS provides valuable information about the electronic structure of TMOs~\cite{veenendaal06,Hariki17,Taguchi16book,taguchi08}. For high-$T_c$ cuprates, the NLS in Cu 2$p_{3/2}$ XPS is extensively used to determine key parameters, such as the CT energy $\Delta_{dp}$, 
and more recently to analyze electronic reconstructions due to doping~\cite{Taguchi05b,Horio18,Okada95,Veenendaal94,taguchi05}.

Further information can be obtained with charge-conserving spectroscopies XAS and RIXS.
The Ni $L_3$-edge 
XAS and RIXS  spectra are measured in both NdNiO$_2$~\cite{Hepting19,Rossi20} and LaNiO$_2$~~\cite{Hepting19}.
Interestingly, a side peak (852.0~eV) is observed in $L_3$-XAS of LaNiO$_2$, while it is absent
in NdNiO$_2$.
A low-energy RIXS feature ($E_{\rm loss}$=0.6~eV) associated with the XAS side peak is observed in LaNiO$_2$. 
The  difference between the Ni $L_3$ XAS and RIXS spectra of NdNiO$_2$ and LaNiO$_2$ poses an open question.

In this paper, we use the local-density approximation (LDA) + dynamical mean-field theory (DMFT)~\cite{metzner89,georges96,kotliar06}  
to calculate XPS, XAS, and RIXS spectra~\cite{Hariki17,Hariki18,Hariki20,Ghiasi19,Hariki20,Jindrich18} 
of undoped infinite-layer nickelates.
By comparison with the available experimental data, we identify the most appropriate
CT energy and use it for classification within
the Zaanen-Sawatzky-Allen scheme~\cite{Zaanen85}.

Material-specific DMFT calculations
for NdNiO$_2$ or LaNiO$_2$ were performed by several authors, leading to contradictory
conclusions, which
can be sorted into two groups:
(i) Multiorbital (Hund's metal) physics is crucial~\cite{Wang20,Kang20,Petocchi20,Lechermann20b}, and (ii) (single-orbital) Mott-Hubbard physics is relevant with little influence of charge-transfer effects or with a small self-doping by Nd 5$d$ electrons~\cite{Karp20,Kitatani20,Karp20b}. 
The differences, recently addressed blue{by Karp, Hampel, and Millis}~\cite{Karp21},
can be traced to the model parameters, which are not uniquely defined, such as the interaction strength, orbital basis, and, in particular, the double-counting correction. 
To settle the debate, an experimental input is needed to provide a benchmark for
selecting the model parameters.

\section{Computational Method}
The XPS, XAS and RIXS simulations start with a standard LDA+DMFT calculation~\cite{georges96,kotliar06,kunes09,Hariki17,Hariki18,Hariki20}. 
First, 
LDA bands for the experimental crystal structure of NdNiO$_2$ and LaNiO$_2$~\cite{Hayward99,li19} are calculated using the Wien2K package~\cite{wien2k}\footnote{The Nd 4$f$ states in NdNiO$_2$ are treated as partially-filled core states.} and
projected onto Wannier basis spanning the Ni 3$d$, O 2$p$, and Nd (La) 5$d$ orbitals~\cite{wien2wannier,wannier90}. 
 The 
 model is augmented with a local electron-electron interaction within the Ni $3d$ shell, parametrized by Coulomb's $U$=5.0~eV and Hund's $J$=1.0~eV~\cite{Kang20,Wang20,Ryee20}.
The strong-coupling continuous-time quantum Monte Carlo impurity solver~\cite{werner06,boehnke11,hafermann12,Hariki15} 
is employed with the DMFT cycle to obtain the  Ni $3d$ self-energy $\Sigma(i\omega_n)$,
which is analytically continued~\cite{jarrell96} to real frequency after
having reached the self-consistency .
The calculations are performed at temperature $T=290$~K.

The XPS, XAS, and RIXS spectra are calculated from the Anderson impurity model
augmented with the 2$p$ core states and the real-frequency hybridization function discretized into 40--50 levels (per spin and orbital). To this end, we use the configuration-interaction solver; for details, see Refs.~\onlinecite{Hariki17,Ghiasi19} for XPS and Refs.~\onlinecite{Hariki18,Hariki20,Winder20} for XAS and RIXS simulation. 

\textcolor{black}{Determination of Ni $3d$ site energies in the model studied by DMFT involves subtracting the so-called double-counting correction $\mu_{\rm{dc}}$ from the respective LDA values ($\varepsilon_d^{\rm LDA}$), a procedure accounting for the effect of the $dd$ interaction present in the LDA description. It is clear that $\mu_{\rm{dc}}$ is of the order of Hartree energy $Un_d$, but a generally accepted universal expression is not available~\cite{kotliar06,karolak10,Haule15}. While a similar uncertainty exists also for interaction parameters $U$ and $J$, impact of their variation on physical properties is usually minor (see Supplemental Material~\cite{SM} for NdNiO$_2$-specific discussion). Variation of $\mu_{\rm{dc}}$, on the other hand, may have a profound effect. Therefore we choose to adjust $\mu_{\rm dc}$ by comparison to the experimental data. Although $\mu_{\rm dc}$ is the parameter entering the calculation, in the discussion we use its linear function $\Delta_{dp}=(\varepsilon_d^{\rm LDA}-\mu_{\rm dc})+9U_{dd}-\varepsilon_p^{\rm LDA}$, which sets the scale for the energy necessary to transfer an electron from O $2p$ to Ni $3d$ orbital. Here, $U_{dd}=U-\tfrac{4}{9}J$ is the average interorbital interaction, and 9 is the Ni $3d$ occupation \textcolor{black}{in the Ni$^+$ formal valence (similar to the definition of the charge-transfer energy in the cluster model~\cite{groot_kotani,Ghiasi19,Hariki20})}. 
}

\begin{figure}[tb]
\includegraphics[width=1.00\columnwidth]{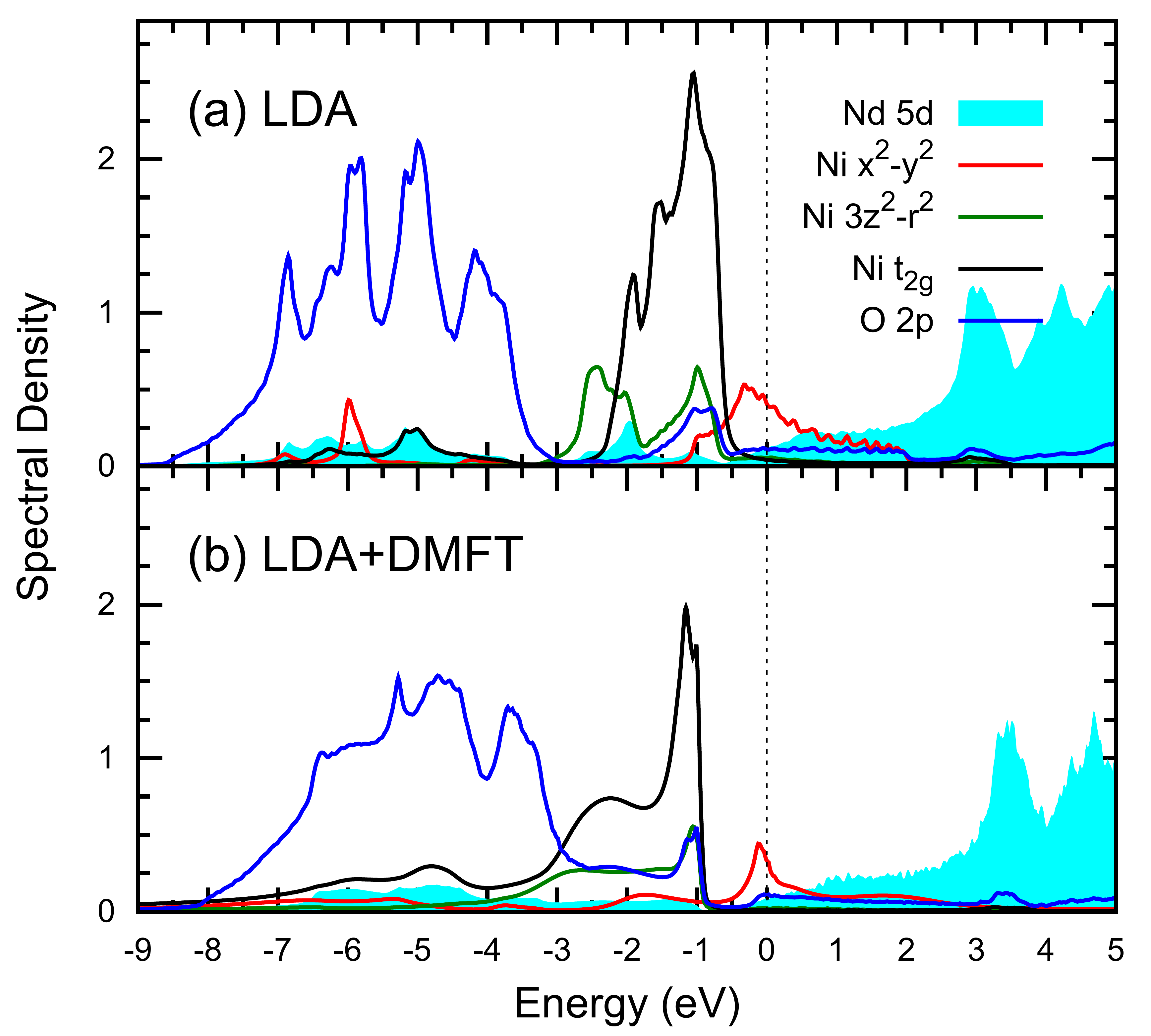}\hspace{0pt}
\caption{The one-particle spectral densities of  NdNiO$_2$ obtained by (a) LDA and (b) LDA+DMFT (for $\Delta_{dp}=4.9$~eV).}
\label{fig:ldadmft}
\end{figure}

\begin{figure}
\includegraphics[width=0.98\columnwidth]{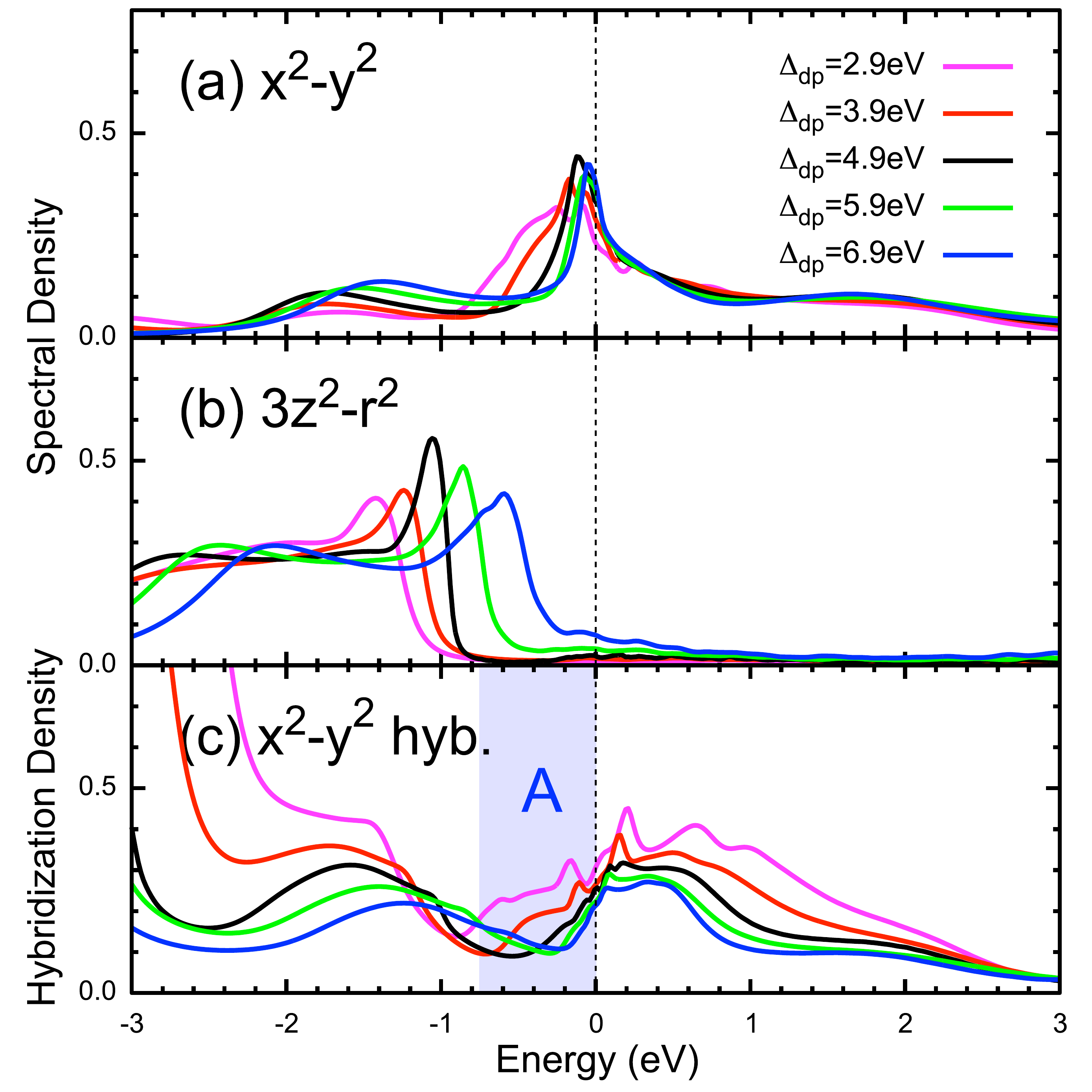}\hspace{0pt}
\caption{The DMFT spectral densities for (a) Ni $x^2-y^2$ and (b) Ni $3z^2-r^2$ orbitals along with (c) the Ni $x^2-y^2$ hybridization function computed for different $\Delta_{dp}$ values.}
\label{fig:doshyb}
\end{figure}

\section{Electronic structure}
\begin{figure}[tb]
\includegraphics[width=1.00\columnwidth]{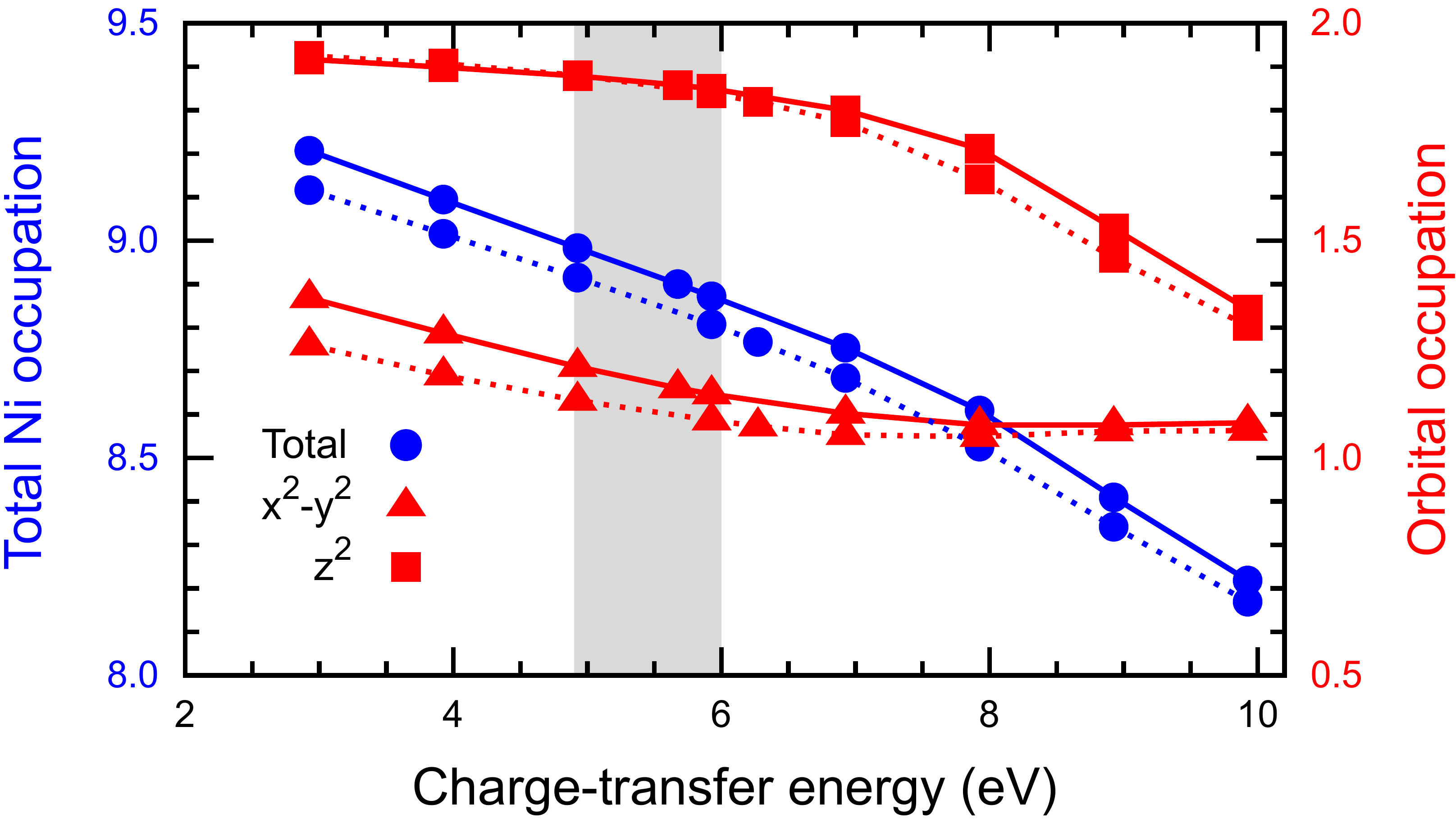}\hspace{0pt}
\caption{The DMFT occupation of $x^2-y^2$ (red, square) and $3z^2-r^2$ (red, triangle) orbitals and the entire Ni $3d$ shell (blue, circle) as a function of $\Delta_{dp}$. The full line is obtained 
for NdNiO$_2$, and the dashed line for Nd$_{0.775}$Sr$_{0.225}$NiO$_2$.}
\label{fig:occ}
\end{figure}

Figure.~\ref{fig:ldadmft} shows the orbitally resolved spectral densities
(projected density of states) of NdNiO$_2$ obtained by LDA and LDA+DMFT for
$\Delta_{dp}=4.9$~eV, which we later identify as the optimal parameter choice. Both the LDA and LDA+DMFT
yield a metallic state with the Ni $x^2-y^2$ orbital character dominating around the Fermi level.
This general picture is valid in the entire range of studied $\Delta_{dp}=2.9$--$6.9$~eV.
In Fig.~\ref{fig:doshyb}, we show the dependence of Ni $x^2-y^2$ and $3z^2-r^2$ spectra
on $\Delta_{dp}$. Increasing $\Delta_{dp}$ corresponds to an upward shift of the bare Ni $3d$ site energies, which is indirectly reflected in the shift of the $3z^2-r^2$ band. The $x^2-y^2$ 
peak at the Fermi level, rather than being shifted, exhibits an increased mass renormalization
(reduced width). The amplitude of the $x^2-y^2$ hybridization function around the Fermi level is
reduced with increasing $\Delta_{dp}$; in particular, the sizable decrease just below
the Fermi level (blue region) has an important implication for the XPS spectra as discussed later.
The evolution of $x^2-y^2$ and $3z^2-r^2$ occupancies in Fig.~\ref{fig:occ} shows that, 
up to $\Delta_{dp}\approx7$~eV the $3z^2-r^2$ is completely filled (the deviation from 
2.0 is due to hybridization with empty bands). The physics is , thus, effectively of a single-orbital
Hubbard model, and the Ni ion takes a monovalent (Ni$^{1+}, d^9$) character.

\begin{figure}[tb]
\includegraphics[width=0.98\columnwidth]{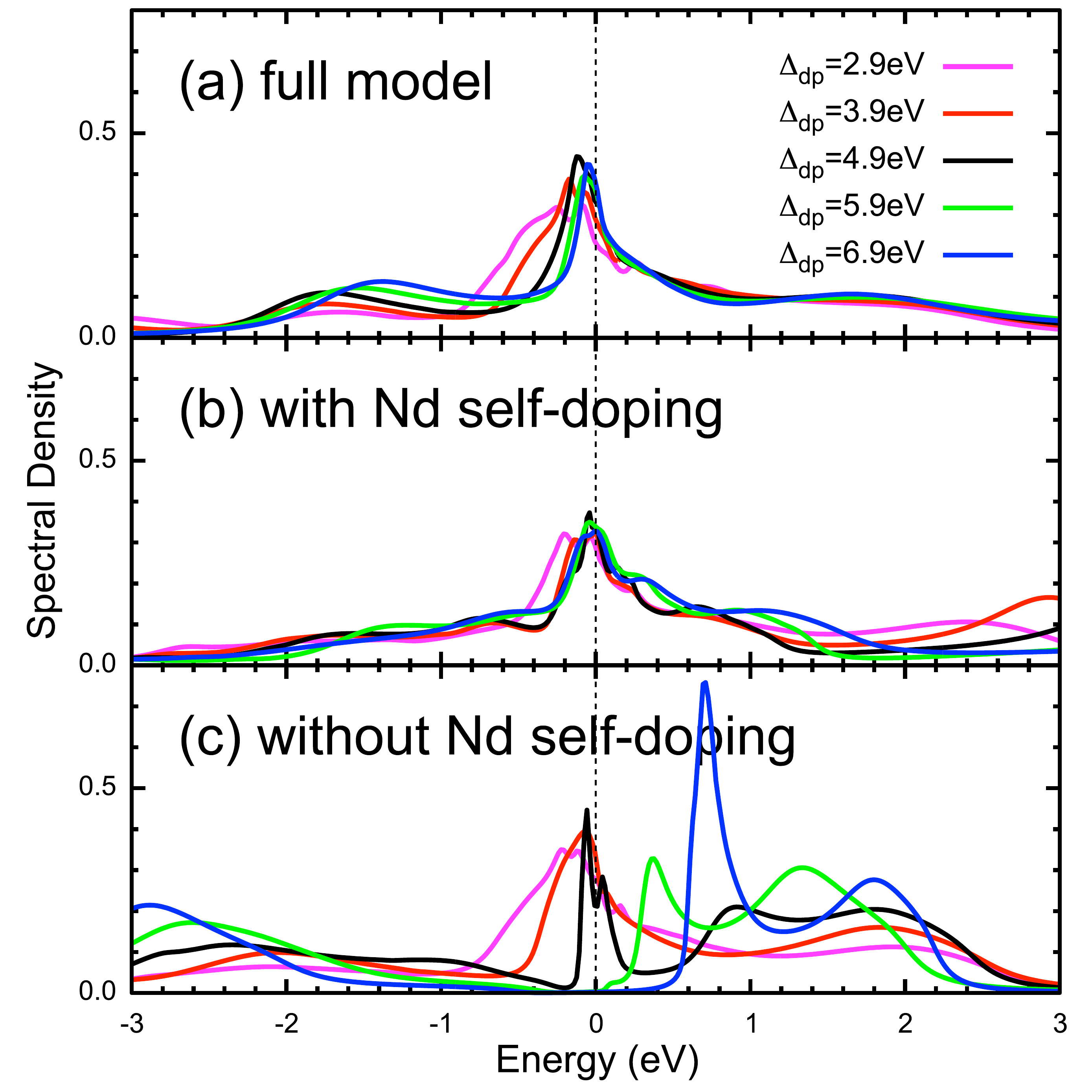}\hspace{0pt}
\caption{The $x^2-y^2$ spectral densities computed in (a) the full model [the same as in Fig.~\ref{fig:doshyb}(a)], (b) model (i) with a self-doping from Nd $d$ bands, and (c) model (ii) without a self-doping from Nd $d$ bands.}
\label{fig:model}
\end{figure}

\textcolor{black}{Different from cuprates, the stoichiometric parent compound is metallic.}
In order to analyze the role of Nd $d$ bands, we study two modified models: (i) hybridization between NiO$_2$ planes and the Nd orbitals is switched off, and (ii) Nd orbitals are removed from the model. In the former case (i) self-doping 
of the NiO$_2$ planes from Nd orbitals is possible, while in the latter case (ii) the stoichiometry of the NiO$_2$ planes cannot change. The evolution of the $x^2-y^2$ spectral density
with $\Delta_{dp}$ for (i) and (ii) is shown in Fig.~\ref{fig:model}. Like the full model, the
low-energy spectrum of model (i) remains metallic over the whole studied range of $\Delta_{dp}$.
Removing the Nd orbitals (ii) results in progressive mass renormalization 
with increasing $\Delta_{dp}$
and eventually
opening of a gap above $\Delta_{dp}=5.9$~eV. This can be understood as a result of effective weakening of the Ni-O hybridization, i.e., a bandwidth-driven Mott transition. 
The NiO$_2$ layers in NdNiO$_2$ can, thus, be viewed as a strongly correlated system in the vicinity 
of Mott transition, where the insulating state is precluded by the presence of Nd $5d$ bands~\cite{Hirayama20}.

\begin{figure}[tb]
\includegraphics[width=0.98\columnwidth]{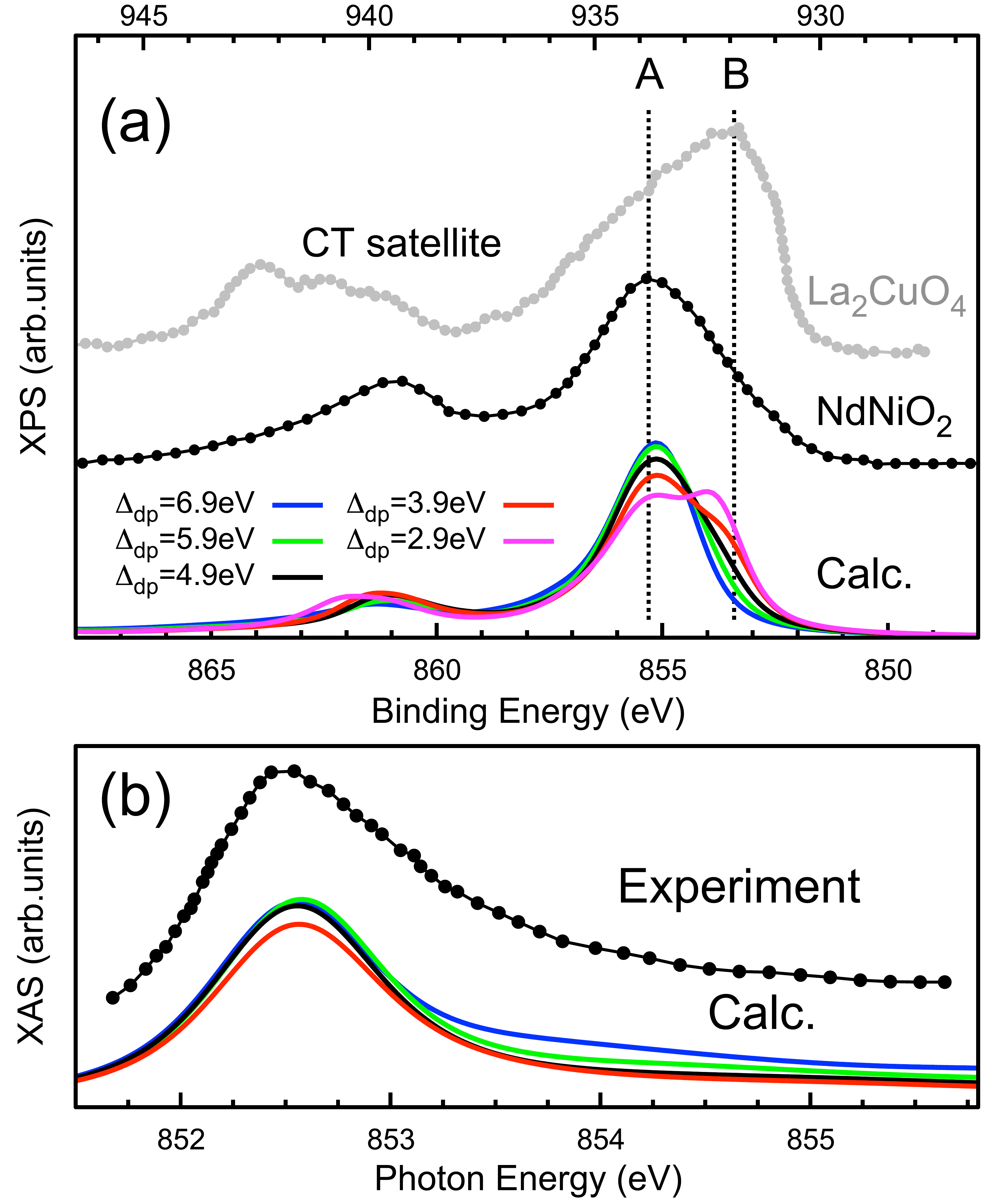}\hspace{0pt}
\caption{(a) Ni 2$p_{3/2}$ XPS spectra and (b)  Ni 2$p_{3/2}$ XAS spectra of NdNiO$_2$ calculated by the LDA+DMFT method for different $\Delta_{dp}$ values. The experimental data~\cite{Fu19,Hepting19} are shown together. For comparison, experimental Cu 2$p_{3/2}$ XPS data of La$_2$CuO$_4$ are shown (gray)~\cite{Taguchi05b}. 
\textcolor{black}{The spectral broadening is taken into account using a Lorentzian 300~meV (HWHM) and a Gaussian 250~meV (HWHM)  for XAS and a Lorentzian 500~meV and a Gaussian 400~meV for XPS. The XPS spectra with different broadening widths can be found in Supplemental Material~\cite{SM}.}
}
\label{fig:ndxpsxas}
\end{figure}

\begin{figure}[tb]
\includegraphics[width=0.98\columnwidth]{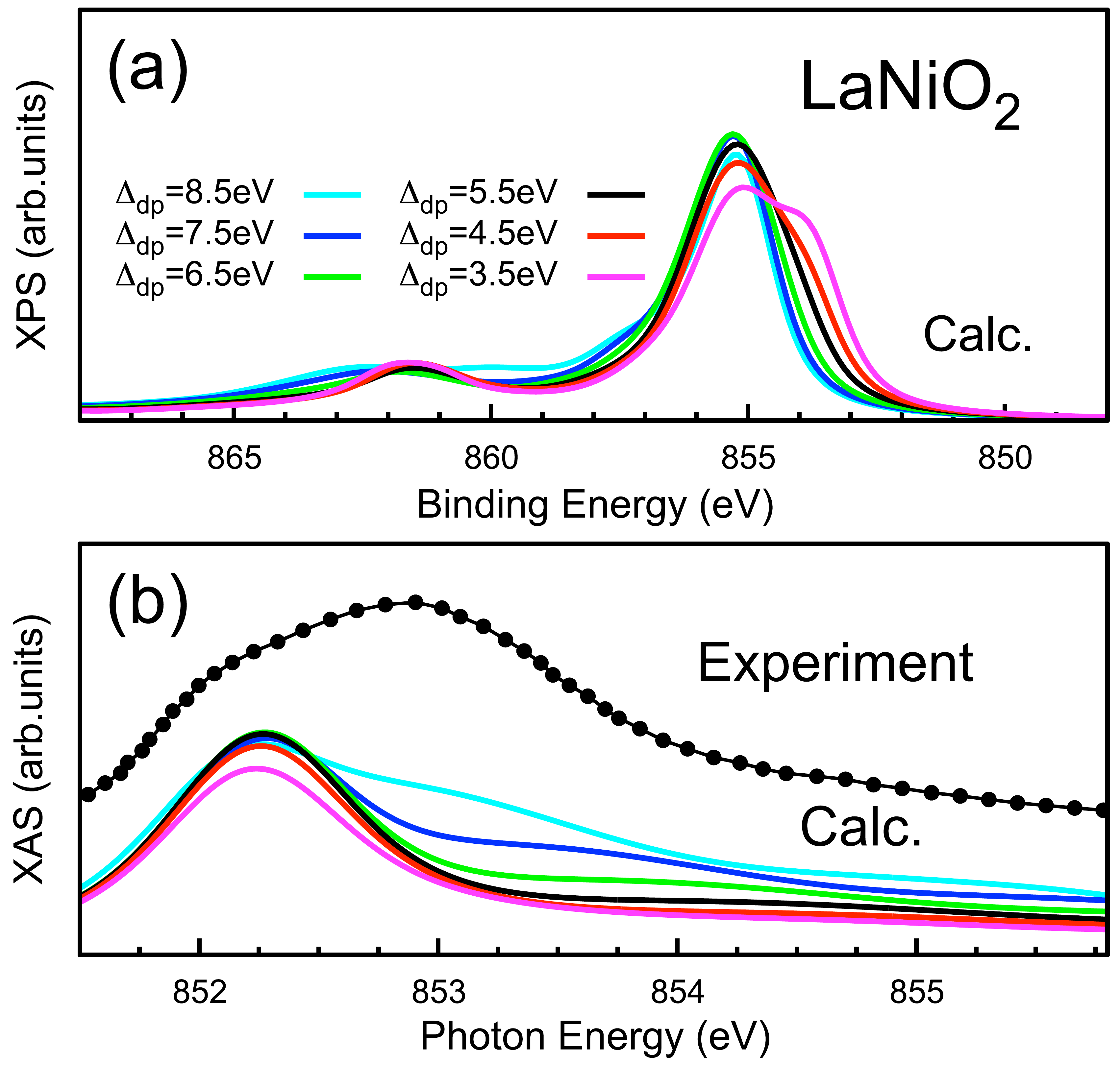}\hspace{0pt}
\caption{(a) Ni 2$p_{3/2}$ XPS spectra and (b)  Ni 2$p_{3/2}$ XAS spectra of LaNiO$_2$ calculated by the LDA+DMFT method for different $\Delta_{dp}$ values.
\textcolor{black}{The spectral broadening is taken into account using a Lorentzian 300~meV (HWHM) and a Gaussian 250~meV (HWHM)  for XAS, and a Lorentzian 500~meV and a Gaussian 400~meV for XPS.}
}
\label{fig:laxpsxas}
\end{figure}


\section{Comparison to experimental x-ray spectroscopies}
\subsection{Ni $2p_{3/2}$ XPS}
Next, we investigate the impact of the variation of $\Delta_{dp}$ on the core-level spectra.
Figure~\ref{fig:ndxpsxas} shows the calculated Ni $2p_{3/2}$ XPS spectra of NdNiO$_2$ together with the experimental data~\cite{Fu19}.
The Ni 2$p_{3/2}$ XPS spectrum consists of two components:~the main-line (852--857~eV) and the CT satellite (861~eV)~\cite{Hariki17,groot_kotani}.
The core hole created by x rays represents an attractive potential, which induces CT from surrounding atoms to the empty 3$d$ orbital on the excited Ni site.  
The main line corresponds to the CT screened final states, while the CT satellite corresponds to unscreened ones~\cite{veenendaal93,taguchi05,Hariki17}.
Fu $et$ $al$.~\cite{Fu19} observe a shoulder $B$ (approximately $856.5$~eV) in the main line. 
Unlike $A$, the peak $B$ is absent in the cluster-model spectra~\cite{Ghiasi19,veenendaal93} and, thus,
can be ascribed to NLS~\cite{Fu19}. The sensitivity of the relative intensity of $A$ and $B$ to $\Delta_{dp}$ can be used to locate its value to the interval $4.9$--$5.9$~eV.
The observed behavior of the NLS feature $B$ reflects the amplitude of the hybridization function
just below the Fermi level~\cite{Hariki17}, the shaded
area in Fig.~\ref{fig:doshyb}(c).

The NLS ($B$) is known to dominate over the local screening ($A$) in cuprates, as
shown in Fig.~\ref{fig:ndxpsxas} for Cu 2$p_{3/2}$ XPS in La$_2$CuO$_4$~\cite{Taguchi05b}. 
For small $\Delta_{dp}=2.9$~eV, a typical value for high-$T_c$ cuprates~\cite{taguchi05,Zaanen85,Taguchi05b,veenendaal93,Ghijsen88},
the spectra of NdNiO$_2$ resemble that of La$_2$CuO$_4$.
Thus our analysis shows that $\Delta_{dp}$ in NdNiO$_2$ is by 2--3~eV larger than in cuprates.
The relative size $\Delta_{dp}$ and the Hubbard $U$ would place NdNiO$_2$ somewhere between the Mott-Hubbard ($\Delta_{dp}>U$) and CT ($\Delta_{dp}<U$) systems in the Zaanen-Sawatzky-Allen classification of TMOs~\cite{Zaanen85,Nomura20,Nomura19,Karp20b}.
\textcolor{black}{The calculated occupations for doped Nd$_{0.775}$Sr$_{0.225}$NiO$_2$, shown in Fig.~\ref{fig:occ} and in Supplementary Material~\cite{SM}, reveal that for optimal $\Delta_{dp}$ doped holes are almost equally shared by Ni, Nd and O sites.
This is a remarkable difference to monovalent cuprates or divalent NiO. In these systems of strong charge-transfer character, the doped holes reside predominantly
in O $2p$ orbitals, irrespective of a substantial $3d$ spectral weight just below the Fermi level~\cite{kunes07a}.}
Moreover, for \textcolor{black}{the optimal $\Delta_{dp}$} values inferred above, the doped holes in NdNiO$_2$  do not enter the Ni $3z^2-r^2$ orbitals (Fig.~\ref{fig:occ}). The 
single-band Hubbard description is thus valid for not only the parent NdNiO$_2$ but also the superconducting one Nd$_{0.8}$Sr$_{0.2}$NiO$_2$, as suggested by Refs.~\cite{Karp20,Kitatani20,Karp20b}.   

\textcolor{black}{Proximity to NiO$_2$ layers to a Mott state (precluded by self-doping from Nd) suggests that a superexchange interaction still plays a role despite the metallic state. Using the optimal $\Delta_{dp}$ we arrive~\cite{SM} at the nearest Ni--Ni anti-ferromagnetic exchange in the range 40--60~meV. 
Given the oversimplification of representing spin response of a metal in terms of local moments interactions, this value is consistent with 69~meV inferred from the RIXS experiment on a related compound La$_4$Ni$_3$O$_8$~\cite{Lin2021}}

The calculated LaNiO$_2$ spectra in Fig.~\ref{fig:laxpsxas}(a) show similar behavior to NdNiO$_2$.

\begin{figure*}[tb]
\includegraphics[width=2.07\columnwidth]{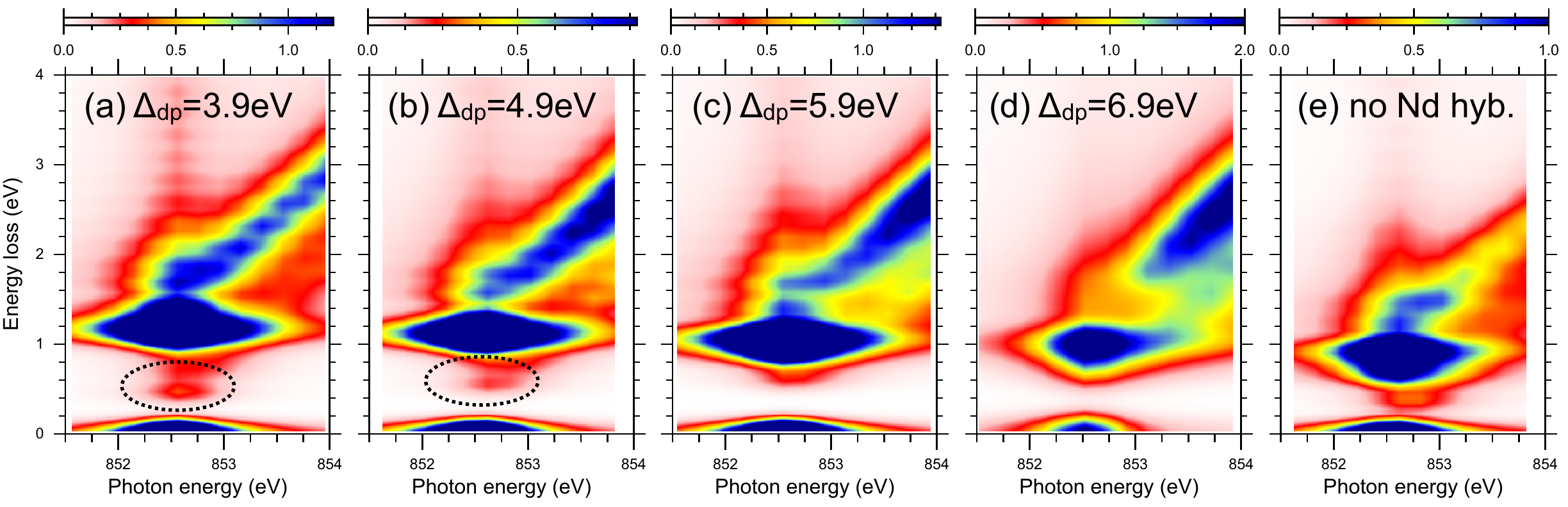}\hspace{0pt}
\caption{The Ni $L_3$ RIXS spectra of NdNiO$_2$ calculated for (a) $\Delta_{dp}=3.9$~eV, (b) $\Delta_{dp}=4.9$~eV, (c) $\Delta_{dp}=5.9$~eV, and (d) $\Delta_{dp}=6.9$~eV. (e) the Ni $L_3$ RIXS spectra calculated for the model without the hybridization between Nd 5$d$ and NiO$_2$ plane ($\Delta_{dp}$=4.9~eV).  The spectral broadening is considered using a Gaussian of 100~meV (HWHM). }
\label{fig:ndrixs}
\end{figure*}

\begin{figure}[tb]
\includegraphics[width=0.98\columnwidth]{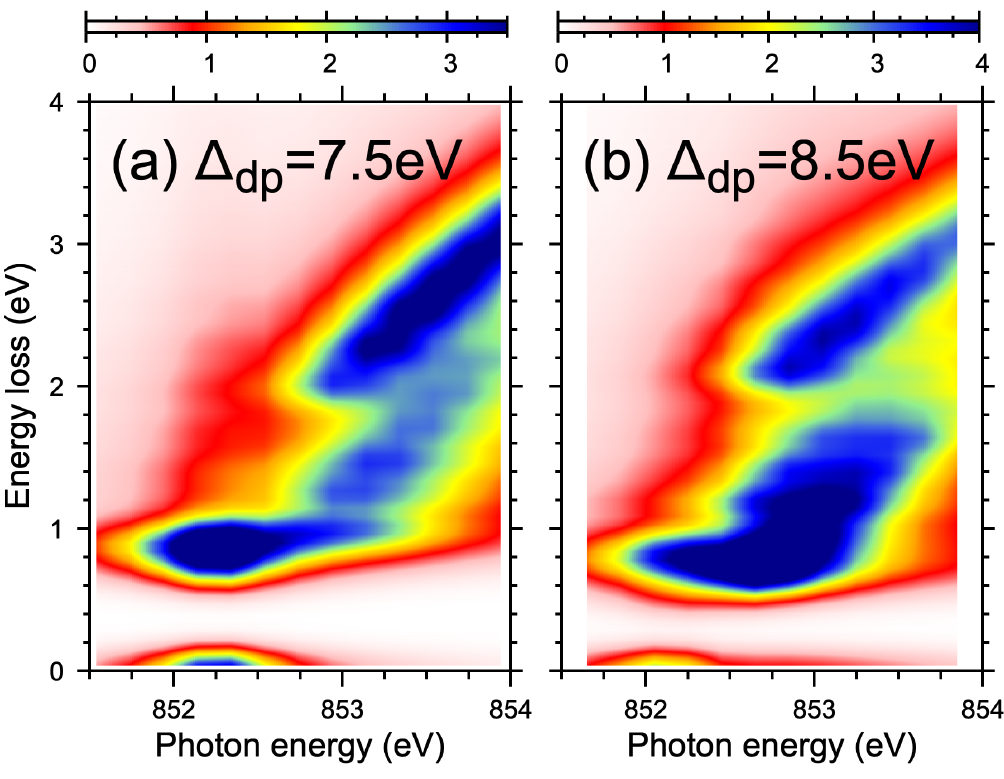}\hspace{0pt}
\caption{The Ni $L_3$ RIXS spectra of LaNiO$_2$ calculated for (a) $\Delta_{dp}=7.5$~eV, (b) $\Delta_{dp}=8.5$~eV. The spectral broadening is considered using a Gaussian of 100~meV (HWHM). }
\label{fig:larixs}
\end{figure}

\subsection{Ni $2p_{3/2}$ XAS and RIXS}
As expected for Ni$^{1+}$ systems with a $d^9$ configuration, the experimental Ni $2p_{3/2}$ XAS of NdNiO$_2$ shows a sharp peak corresponding to the electron excitation from the 2$p_{3/2}$ to an empty $x^2-y^2$ orbital [Fig.~\ref{fig:ndxpsxas}(b)]. 
The XAS main peak is accompanied by a broad tail attributed to the hybridization with metallic bands. 
The theoretical results in Fig.~\ref{fig:ndxpsxas}(b) reproduce the experimental data reasonably well;~however, the weak dependence
on $\Delta_{dp}$ does not allow to draw conclusions about its value.

The RIXS spectra, on the other hand, exhibit fine changes with the $\Delta_{dp}$ values, see Fig.~\ref{fig:ndrixs}. The spectra at all $\Delta_{dp}$ values contain a strong Raman-like (RL) feature (at constant $E_{\rm loss}$ irrespective of the incident photon energies $E_{\rm in}$) at $E_{\rm loss}\sim$1~eV and a fluorescence-like (FL) feature ($E_{\rm loss}$ linearly increases with $E_{\rm in}$). The RL feature arises from $t_{2g}\rightarrow x^2-y^2$ excitation, and its width (in $E_{\rm loss}$) reflects a rapid decay of this local "exciton". With increasing
$\Delta_{dp}$, the RL feature shifts to lower energies, due upward shift to the $t_{2g}$ bands similar to
$3z^2-r^2$ shown in Fig.~\ref{fig:doshyb}(b), while the $x^2-y^2$ peaks remain pinned in the vicinity of the Fermi level. The main variation of the RIXS spectra with increasing $\Delta_{dp}$ concerns the behavior of the FL part, the onset of which is pushed to higher $E_{\rm loss}$. 
For $\Delta_{dp}$=4.9~eV, deduced from the XPS data, the FL feature sets in below the RL feature at around $E_{\rm loss}\sim$0.6~eV. 
The coexisting RL and FL features above well capture the experimental data by Hepting $et~al$.~\cite{Hepting19} and Rossi $et~al$.~\cite{Rossi20}.
\textcolor{black}{Artificial suppression of hybridization to Nd $5d$ states [Fig.~\ref{fig:ndrixs}(e)] leads to a reduced
intensity of the FL feature and only a moderate modification of the low-energy spectra supporting
the conclusion about the electron-reservoir role of Nd $5d$ states.}

Finally we discuss XAS and RIXS spectra in LaNiO$_2$ (the experimental XPS data are not available
at the moment). The experimental XAS spectra of LaNiO$_2$~\cite{Hepting19} are clearly distinct from NdNiO$_2$.
A side peak at 852.0~eV is attributed to Ni--La hybridization effect by Hepting $et~al$~\cite{Hepting19} based on a simplified impurity model simulation.
\textcolor{black}{The LDA+DMFT calculations (including Ni--La hybridization) do not support this conclusion as they do not match the experimental XAS spectra. While large $\Delta_{dp}$ gives rise to a high-energy XAS shoulder (Fig.~\ref{fig:laxpsxas}), it does not improve the agreement of the RIXS spectra, shown in Fig.~\ref{fig:larixs}. We have to conclude that the present LDA+DMFT description of LaNiO$_2$ does not match the experiment for any choice of $\Delta_{dp}$.}

\textcolor{black}{We propose that the problem lies on the experimental side;~i.e., the measured spectra do not represent a perfect LaNiO$_2$ crystal. We argue by the success of the present method for a broad spectrum of transition-metal oxides~\cite{Hariki17}
including NdNiO$_2$ as well as the absence of an  obvious source of difference between NdNiO$_2$ and LaNiO$_2$. On the experimental side, we point out recent studies~\cite{Zeng21,Osada21} reporting superconductivity in Sr-doped LaNiO$_2$, suggesting that NdNiO$_2$ and LaNiO$_2$ are not that different after all. Spectroscopic experiments on these new LaNiO$_2$ samples are needed to resolve the present discrepancy.} 



\section{\label{sec:sum}Conclusions}
We have presented a comprehensive analysis of Ni 2$p_{3/2}$ core-level XPS, XAS, and RIXS in infinite-layer nickelates (NdNiO$_2$ and LaNiO$_2$) with the LDA+DMFT approach. Comparison to the experimental spectra allowed us
to determine the CT parameter (double-counting correction) and make the following conclusions about the electronic
structure. Undoped NdNiO$_2$ is nearly monovalent (Ni$^{1+}, d^9$) with a small self-doping from the Nd 5$d$ band.
Only the Ni $x^2-y^2$ orbitals are partially filled and multiorbital physics does not play an important role
for the stoichiometric as well as slightly hole-doped compound. Unlike in cuprates, the Ni-O hybridization does 
not play an important role in connection with doping -- doped holes reside predominantly on the Ni sites.
The physics of NdNiO$_2$ described effectively by a 
single-band Hubbard model~\cite{Karp20,Kitatani20,Karp20b}
is consistent with the available core-level spectroscopies.
While the present calculations provide a good description of the experimental core-level spectra of NdNiO$_2$,
we cannot explain the qualitative difference between the reported NdNiO$_2$ and LaNiO$_2$ XAS and RIXS spectra.



\begin{acknowledgments}

We thank M. Kitatani, K. Yamagami, T. Uozumi, H. Ikeno, L. Si, M.-J. Huang and R.-P. Wang for valued discussions.
A.H., M.W., and J.K. were supported by the European Research Council (ERC) under the European Union's Horizon 2020 research and innovation program (Grant Agreement No.~646807-EXMAG). A.H. was supported by JSPS KAKENHI Grant No. 21K13884. The numerical calculations were performed at the Vienna Scientific Cluster (VSC).

\end{acknowledgments}

\bibliography{main}

\begin{thebibliography}{63}%
\makeatletter
\providecommand \@ifxundefined [1]{%
 \@ifx{#1\undefined}
}%
\providecommand \@ifnum [1]{%
 \ifnum #1\expandafter \@firstoftwo
 \else \expandafter \@secondoftwo
 \fi
}%
\providecommand \@ifx [1]{%
 \ifx #1\expandafter \@firstoftwo
 \else \expandafter \@secondoftwo
 \fi
}%
\providecommand \natexlab [1]{#1}%
\providecommand \enquote  [1]{``#1''}%
\providecommand \bibnamefont  [1]{#1}%
\providecommand \bibfnamefont [1]{#1}%
\providecommand \citenamefont [1]{#1}%
\providecommand \href@noop [0]{\@secondoftwo}%
\providecommand \href [0]{\begingroup \@sanitize@url \@href}%
\providecommand \@href[1]{\@@startlink{#1}\@@href}%
\providecommand \@@href[1]{\endgroup#1\@@endlink}%
\providecommand \@sanitize@url [0]{\catcode `\\12\catcode `\$12\catcode
  `\&12\catcode `\#12\catcode `\^12\catcode `\_12\catcode `\%12\relax}%
\providecommand \@@startlink[1]{}%
\providecommand \@@endlink[0]{}%
\providecommand \url  [0]{\begingroup\@sanitize@url \@url }%
\providecommand \@url [1]{\endgroup\@href {#1}{\urlprefix }}%
\providecommand \urlprefix  [0]{URL }%
\providecommand \Eprint [0]{\href }%
\providecommand \doibase [0]{http://dx.doi.org/}%
\providecommand \selectlanguage [0]{\@gobble}%
\providecommand \bibinfo  [0]{\@secondoftwo}%
\providecommand \bibfield  [0]{\@secondoftwo}%
\providecommand \translation [1]{[#1]}%
\providecommand \BibitemOpen [0]{}%
\providecommand \bibitemStop [0]{}%
\providecommand \bibitemNoStop [0]{.\EOS\space}%
\providecommand \EOS [0]{\spacefactor3000\relax}%
\providecommand \BibitemShut  [1]{\csname bibitem#1\endcsname}%
\let\auto@bib@innerbib\@empty
\bibitem [{\citenamefont {Bednorz}\ and\ \citenamefont
  {M{\"u}ller}(1986)}]{Bednorz86}%
  \BibitemOpen
  \bibfield  {author} {\bibinfo {author} {\bibfnamefont {J.~G.}\ \bibnamefont
  {Bednorz}}\ and\ \bibinfo {author} {\bibfnamefont {K.~A.}\ \bibnamefont
  {M{\"u}ller}},\ }\href {\doibase 10.1007/BF01303701} {\bibfield  {journal}
  {\bibinfo  {journal} {Z. Phys., B Condens. matter}\ }\textbf {\bibinfo
  {volume} {64}},\ \bibinfo {pages} {189} (\bibinfo {year} {1986})}\BibitemShut
  {NoStop}%
\bibitem [{\citenamefont {Imada}\ \emph {et~al.}(1998)\citenamefont {Imada},
  \citenamefont {Fujimori},\ and\ \citenamefont {Tokura}}]{imada98}%
  \BibitemOpen
  \bibfield  {author} {\bibinfo {author} {\bibfnamefont {M.}~\bibnamefont
  {Imada}}, \bibinfo {author} {\bibfnamefont {A.}~\bibnamefont {Fujimori}}, \
  and\ \bibinfo {author} {\bibfnamefont {Y.}~\bibnamefont {Tokura}},\ }\href
  {\doibase 10.1103/RevModPhys.70.1039} {\bibfield  {journal} {\bibinfo
  {journal} {Rev. Mod. Phys.}\ }\textbf {\bibinfo {volume} {70}},\ \bibinfo
  {pages} {1039} (\bibinfo {year} {1998})}\BibitemShut {NoStop}%
\bibitem [{\citenamefont {Dagotto}(1994)}]{Dagotto94}%
  \BibitemOpen
  \bibfield  {author} {\bibinfo {author} {\bibfnamefont {E.}~\bibnamefont
  {Dagotto}},\ }\href {\doibase 10.1103/RevModPhys.66.763} {\bibfield
  {journal} {\bibinfo  {journal} {Rev. Mod. Phys.}\ }\textbf {\bibinfo {volume}
  {66}},\ \bibinfo {pages} {763} (\bibinfo {year} {1994})}\BibitemShut
  {NoStop}%
\bibitem [{\citenamefont {Li}\ \emph {et~al.}(2019)\citenamefont {Li},
  \citenamefont {Lee}, \citenamefont {Wang}, \citenamefont {Osada},
  \citenamefont {Crossley}, \citenamefont {Lee}, \citenamefont {Cui},
  \citenamefont {Hikita},\ and\ \citenamefont {Hwang}}]{li19}%
  \BibitemOpen
  \bibfield  {author} {\bibinfo {author} {\bibfnamefont {D.}~\bibnamefont
  {Li}}, \bibinfo {author} {\bibfnamefont {K.}~\bibnamefont {Lee}}, \bibinfo
  {author} {\bibfnamefont {B.~Y.}\ \bibnamefont {Wang}}, \bibinfo {author}
  {\bibfnamefont {M.}~\bibnamefont {Osada}}, \bibinfo {author} {\bibfnamefont
  {S.}~\bibnamefont {Crossley}}, \bibinfo {author} {\bibfnamefont {H.~R.}\
  \bibnamefont {Lee}}, \bibinfo {author} {\bibfnamefont {Y.}~\bibnamefont
  {Cui}}, \bibinfo {author} {\bibfnamefont {Y.}~\bibnamefont {Hikita}}, \ and\
  \bibinfo {author} {\bibfnamefont {H.~Y.}\ \bibnamefont {Hwang}},\ }\href
  {\doibase 10.1038/s41586-019-1496-5} {\bibfield  {journal} {\bibinfo
  {journal} {Nature}\ }\textbf {\bibinfo {volume} {572}},\ \bibinfo {pages}
  {624} (\bibinfo {year} {2019})}\BibitemShut {NoStop}%
\bibitem [{\citenamefont {Botana}\ and\ \citenamefont
  {Norman}(2020)}]{Botana20}%
  \BibitemOpen
  \bibfield  {author} {\bibinfo {author} {\bibfnamefont {A.~S.}\ \bibnamefont
  {Botana}}\ and\ \bibinfo {author} {\bibfnamefont {M.~R.}\ \bibnamefont
  {Norman}},\ }\href {\doibase 10.1103/PhysRevX.10.011024} {\bibfield
  {journal} {\bibinfo  {journal} {Phys. Rev. X}\ }\textbf {\bibinfo {volume}
  {10}},\ \bibinfo {pages} {011024} (\bibinfo {year} {2020})}\BibitemShut
  {NoStop}%
\bibitem [{\citenamefont {Krishna}\ \emph {et~al.}(2020)\citenamefont
  {Krishna}, \citenamefont {LaBollita}, \citenamefont {Fumega}, \citenamefont
  {Pardo},\ and\ \citenamefont {Botana}}]{Krishna20}%
  \BibitemOpen
  \bibfield  {author} {\bibinfo {author} {\bibfnamefont {J.}~\bibnamefont
  {Krishna}}, \bibinfo {author} {\bibfnamefont {H.}~\bibnamefont {LaBollita}},
  \bibinfo {author} {\bibfnamefont {A.~O.}\ \bibnamefont {Fumega}}, \bibinfo
  {author} {\bibfnamefont {V.}~\bibnamefont {Pardo}}, \ and\ \bibinfo {author}
  {\bibfnamefont {A.~S.}\ \bibnamefont {Botana}},\ }\href {\doibase
  10.1103/PhysRevB.102.224506} {\bibfield  {journal} {\bibinfo  {journal}
  {Phys. Rev. B}\ }\textbf {\bibinfo {volume} {102}},\ \bibinfo {pages}
  {224506} (\bibinfo {year} {2020})}\BibitemShut {NoStop}%
\bibitem [{\citenamefont {Lee}\ and\ \citenamefont {Pickett}(2004)}]{Lee04}%
  \BibitemOpen
  \bibfield  {author} {\bibinfo {author} {\bibfnamefont {K.-W.}\ \bibnamefont
  {Lee}}\ and\ \bibinfo {author} {\bibfnamefont {W.~E.}\ \bibnamefont
  {Pickett}},\ }\href {\doibase 10.1103/PhysRevB.70.165109} {\bibfield
  {journal} {\bibinfo  {journal} {Phys. Rev. B}\ }\textbf {\bibinfo {volume}
  {70}},\ \bibinfo {pages} {165109} (\bibinfo {year} {2004})}\BibitemShut
  {NoStop}%
\bibitem [{\citenamefont {Zhang}\ \emph {et~al.}(2020)\citenamefont {Zhang},
  \citenamefont {Yang},\ and\ \citenamefont {Zhang}}]{Zhang20}%
  \BibitemOpen
  \bibfield  {author} {\bibinfo {author} {\bibfnamefont {G.-M.}\ \bibnamefont
  {Zhang}}, \bibinfo {author} {\bibfnamefont {Y.-f.}\ \bibnamefont {Yang}}, \
  and\ \bibinfo {author} {\bibfnamefont {F.-C.}\ \bibnamefont {Zhang}},\ }\href
  {\doibase 10.1103/PhysRevB.101.020501} {\bibfield  {journal} {\bibinfo
  {journal} {Phys. Rev. B}\ }\textbf {\bibinfo {volume} {101}},\ \bibinfo
  {pages} {020501} (\bibinfo {year} {2020})}\BibitemShut {NoStop}%
\bibitem [{\citenamefont {Hepting}\ \emph {et~al.}(2020)\citenamefont
  {Hepting}, \citenamefont {Li}, \citenamefont {Jia}, \citenamefont {Lu},
  \citenamefont {Paris}, \citenamefont {Tseng}, \citenamefont {Feng},
  \citenamefont {Osada}, \citenamefont {Been}, \citenamefont {Hikita},
  \citenamefont {Chuang}, \citenamefont {Hussain}, \citenamefont {Zhou},
  \citenamefont {Nag}, \citenamefont {Garcia-Fernandez}, \citenamefont {Rossi},
  \citenamefont {Huang}, \citenamefont {Huang}, \citenamefont {Shen},
  \citenamefont {Schmitt}, \citenamefont {Hwang}, \citenamefont {Moritz},
  \citenamefont {Zaanen}, \citenamefont {Devereaux},\ and\ \citenamefont
  {Lee}}]{Hepting19}%
  \BibitemOpen
  \bibfield  {author} {\bibinfo {author} {\bibfnamefont {M.}~\bibnamefont
  {Hepting}}, \bibinfo {author} {\bibfnamefont {D.}~\bibnamefont {Li}},
  \bibinfo {author} {\bibfnamefont {C.~J.}\ \bibnamefont {Jia}}, \bibinfo
  {author} {\bibfnamefont {H.}~\bibnamefont {Lu}}, \bibinfo {author}
  {\bibfnamefont {E.}~\bibnamefont {Paris}}, \bibinfo {author} {\bibfnamefont
  {Y.}~\bibnamefont {Tseng}}, \bibinfo {author} {\bibfnamefont
  {X.}~\bibnamefont {Feng}}, \bibinfo {author} {\bibfnamefont {M.}~\bibnamefont
  {Osada}}, \bibinfo {author} {\bibfnamefont {E.}~\bibnamefont {Been}},
  \bibinfo {author} {\bibfnamefont {Y.}~\bibnamefont {Hikita}}, \bibinfo
  {author} {\bibfnamefont {Y.-D.}\ \bibnamefont {Chuang}}, \bibinfo {author}
  {\bibfnamefont {Z.}~\bibnamefont {Hussain}}, \bibinfo {author} {\bibfnamefont
  {K.~J.}\ \bibnamefont {Zhou}}, \bibinfo {author} {\bibfnamefont
  {A.}~\bibnamefont {Nag}}, \bibinfo {author} {\bibfnamefont {M.}~\bibnamefont
  {Garcia-Fernandez}}, \bibinfo {author} {\bibfnamefont {M.}~\bibnamefont
  {Rossi}}, \bibinfo {author} {\bibfnamefont {H.~Y.}\ \bibnamefont {Huang}},
  \bibinfo {author} {\bibfnamefont {D.~J.}\ \bibnamefont {Huang}}, \bibinfo
  {author} {\bibfnamefont {Z.~X.}\ \bibnamefont {Shen}}, \bibinfo {author}
  {\bibfnamefont {T.}~\bibnamefont {Schmitt}}, \bibinfo {author} {\bibfnamefont
  {H.~Y.}\ \bibnamefont {Hwang}}, \bibinfo {author} {\bibfnamefont
  {B.}~\bibnamefont {Moritz}}, \bibinfo {author} {\bibfnamefont
  {J.}~\bibnamefont {Zaanen}}, \bibinfo {author} {\bibfnamefont {T.~P.}\
  \bibnamefont {Devereaux}}, \ and\ \bibinfo {author} {\bibfnamefont {W.~S.}\
  \bibnamefont {Lee}},\ }\href {\doibase 10.1038/s41563-019-0585-z} {\bibfield
  {journal} {\bibinfo  {journal} {Nat. Mater.}\ }\textbf {\bibinfo {volume}
  {19}},\ \bibinfo {pages} {381} (\bibinfo {year} {2020})}\BibitemShut
  {NoStop}%
\bibitem [{\citenamefont {Zaanen}\ \emph {et~al.}(1985)\citenamefont {Zaanen},
  \citenamefont {Sawatzky},\ and\ \citenamefont {Allen}}]{Zaanen85}%
  \BibitemOpen
  \bibfield  {author} {\bibinfo {author} {\bibfnamefont {J.}~\bibnamefont
  {Zaanen}}, \bibinfo {author} {\bibfnamefont {G.~A.}\ \bibnamefont
  {Sawatzky}}, \ and\ \bibinfo {author} {\bibfnamefont {J.~W.}\ \bibnamefont
  {Allen}},\ }\href {\doibase 10.1103/PhysRevLett.55.418} {\bibfield  {journal}
  {\bibinfo  {journal} {Phys. Rev. Lett.}\ }\textbf {\bibinfo {volume} {55}},\
  \bibinfo {pages} {418} (\bibinfo {year} {1985})}\BibitemShut {NoStop}%
\bibitem [{\citenamefont {Kune\ifmmode~\check{s}\else \v{s}\fi{}}\ \emph
  {et~al.}(2007{\natexlab{a}})\citenamefont {Kune\ifmmode~\check{s}\else
  \v{s}\fi{}}, \citenamefont {Anisimov}, \citenamefont {Skornyakov},
  \citenamefont {Lukoyanov},\ and\ \citenamefont {Vollhardt}}]{kunes07b}%
  \BibitemOpen
  \bibfield  {author} {\bibinfo {author} {\bibfnamefont {J.}~\bibnamefont
  {Kune\ifmmode~\check{s}\else \v{s}\fi{}}}, \bibinfo {author} {\bibfnamefont
  {V.~I.}\ \bibnamefont {Anisimov}}, \bibinfo {author} {\bibfnamefont {S.~L.}\
  \bibnamefont {Skornyakov}}, \bibinfo {author} {\bibfnamefont {A.~V.}\
  \bibnamefont {Lukoyanov}}, \ and\ \bibinfo {author} {\bibfnamefont
  {D.}~\bibnamefont {Vollhardt}},\ }\href {\doibase
  10.1103/PhysRevLett.99.156404} {\bibfield  {journal} {\bibinfo  {journal}
  {Phys. Rev. Lett.}\ }\textbf {\bibinfo {volume} {99}},\ \bibinfo {pages}
  {156404} (\bibinfo {year} {2007}{\natexlab{a}})}\BibitemShut {NoStop}%
\bibitem [{\citenamefont {{Fu}}\ \emph {et~al.}(2019)\citenamefont {{Fu}},
  \citenamefont {{Wang}}, \citenamefont {{Cheng}}, \citenamefont {{Pei}},
  \citenamefont {{Zhou}}, \citenamefont {{Chen}}, \citenamefont {{Wang}},
  \citenamefont {{Zhao}}, \citenamefont {{Jiang}}, \citenamefont {{Liu}},
  \citenamefont {{Huang}}, \citenamefont {{Wang}}, \citenamefont {{Zhao}},
  \citenamefont {{Yu}}, \citenamefont {{Ye}}, \citenamefont {{Wang}},\ and\
  \citenamefont {{Mei}}}]{Fu19}%
  \BibitemOpen
  \bibfield  {author} {\bibinfo {author} {\bibfnamefont {Y.}~\bibnamefont
  {{Fu}}}, \bibinfo {author} {\bibfnamefont {L.}~\bibnamefont {{Wang}}},
  \bibinfo {author} {\bibfnamefont {H.}~\bibnamefont {{Cheng}}}, \bibinfo
  {author} {\bibfnamefont {S.}~\bibnamefont {{Pei}}}, \bibinfo {author}
  {\bibfnamefont {X.}~\bibnamefont {{Zhou}}}, \bibinfo {author} {\bibfnamefont
  {J.}~\bibnamefont {{Chen}}}, \bibinfo {author} {\bibfnamefont
  {S.}~\bibnamefont {{Wang}}}, \bibinfo {author} {\bibfnamefont
  {R.}~\bibnamefont {{Zhao}}}, \bibinfo {author} {\bibfnamefont
  {W.}~\bibnamefont {{Jiang}}}, \bibinfo {author} {\bibfnamefont
  {C.}~\bibnamefont {{Liu}}}, \bibinfo {author} {\bibfnamefont
  {M.}~\bibnamefont {{Huang}}}, \bibinfo {author} {\bibfnamefont
  {X.}~\bibnamefont {{Wang}}}, \bibinfo {author} {\bibfnamefont
  {Y.}~\bibnamefont {{Zhao}}}, \bibinfo {author} {\bibfnamefont
  {D.}~\bibnamefont {{Yu}}}, \bibinfo {author} {\bibfnamefont {F.}~\bibnamefont
  {{Ye}}}, \bibinfo {author} {\bibfnamefont {S.}~\bibnamefont {{Wang}}}, \ and\
  \bibinfo {author} {\bibfnamefont {J.-W.}\ \bibnamefont {{Mei}}},\ }\href@noop
  {} {\  (\bibinfo {year} {2019})},\ \Eprint {http://arxiv.org/abs/1911.03177}
  {arXiv:1911.03177} \BibitemShut {NoStop}%
\bibitem [{\citenamefont {{Rossi}}\ \emph {et~al.}(2020)\citenamefont
  {{Rossi}}, \citenamefont {{Lu}}, \citenamefont {{Nag}}, \citenamefont {{Li}},
  \citenamefont {{Osada}}, \citenamefont {{Lee}}, \citenamefont {{Wang}},
  \citenamefont {{Agrestini}}, \citenamefont {{Garcia-Fernand ez}},
  \citenamefont {{Chuang}}, \citenamefont {{Shen}}, \citenamefont {{Hwang}},
  \citenamefont {{Moritz}}, \citenamefont {{Zhou}}, \citenamefont
  {{Devereaux}},\ and\ \citenamefont {{Lee}}}]{Rossi20}%
  \BibitemOpen
  \bibfield  {author} {\bibinfo {author} {\bibfnamefont {M.}~\bibnamefont
  {{Rossi}}}, \bibinfo {author} {\bibfnamefont {H.}~\bibnamefont {{Lu}}},
  \bibinfo {author} {\bibfnamefont {A.}~\bibnamefont {{Nag}}}, \bibinfo
  {author} {\bibfnamefont {D.}~\bibnamefont {{Li}}}, \bibinfo {author}
  {\bibfnamefont {M.}~\bibnamefont {{Osada}}}, \bibinfo {author} {\bibfnamefont
  {K.}~\bibnamefont {{Lee}}}, \bibinfo {author} {\bibfnamefont {B.~Y.}\
  \bibnamefont {{Wang}}}, \bibinfo {author} {\bibfnamefont {S.}~\bibnamefont
  {{Agrestini}}}, \bibinfo {author} {\bibfnamefont {M.}~\bibnamefont
  {{Garcia-Fernand ez}}}, \bibinfo {author} {\bibfnamefont {Y.~D.}\
  \bibnamefont {{Chuang}}}, \bibinfo {author} {\bibfnamefont {Z.~X.}\
  \bibnamefont {{Shen}}}, \bibinfo {author} {\bibfnamefont {H.~Y.}\
  \bibnamefont {{Hwang}}}, \bibinfo {author} {\bibfnamefont {B.}~\bibnamefont
  {{Moritz}}}, \bibinfo {author} {\bibfnamefont {K.-J.}\ \bibnamefont
  {{Zhou}}}, \bibinfo {author} {\bibfnamefont {T.~P.}\ \bibnamefont
  {{Devereaux}}}, \ and\ \bibinfo {author} {\bibfnamefont {W.~S.}\ \bibnamefont
  {{Lee}}},\ }\href@noop {} {\  (\bibinfo {year} {2020})},\ \Eprint
  {http://arxiv.org/abs/2011.00595} {arXiv:2011.00595} \BibitemShut {NoStop}%
\bibitem [{\citenamefont {van Veenendaal}\ and\ \citenamefont
  {Sawatzky}(1993)}]{veenendaal93}%
  \BibitemOpen
  \bibfield  {author} {\bibinfo {author} {\bibfnamefont {M.~A.}\ \bibnamefont
  {van Veenendaal}}\ and\ \bibinfo {author} {\bibfnamefont {G.~A.}\
  \bibnamefont {Sawatzky}},\ }\href {\doibase 10.1103/PhysRevLett.70.2459}
  {\bibfield  {journal} {\bibinfo  {journal} {Phys. Rev. Lett.}\ }\textbf
  {\bibinfo {volume} {70}},\ \bibinfo {pages} {2459} (\bibinfo {year}
  {1993})}\BibitemShut {NoStop}%
\bibitem [{\citenamefont {van Veenendaal}(2006)}]{veenendaal06}%
  \BibitemOpen
  \bibfield  {author} {\bibinfo {author} {\bibfnamefont {M.}~\bibnamefont {van
  Veenendaal}},\ }\href {\doibase 10.1103/PhysRevB.74.085118} {\bibfield
  {journal} {\bibinfo  {journal} {Phys. Rev. B}\ }\textbf {\bibinfo {volume}
  {74}},\ \bibinfo {pages} {085118} (\bibinfo {year} {2006})}\BibitemShut
  {NoStop}%
\bibitem [{\citenamefont {Hariki}\ \emph {et~al.}(2017)\citenamefont {Hariki},
  \citenamefont {Uozumi},\ and\ \citenamefont {Kune\ifmmode~\check{s}\else
  \v{s}\fi{}}}]{Hariki17}%
  \BibitemOpen
  \bibfield  {author} {\bibinfo {author} {\bibfnamefont {A.}~\bibnamefont
  {Hariki}}, \bibinfo {author} {\bibfnamefont {T.}~\bibnamefont {Uozumi}}, \
  and\ \bibinfo {author} {\bibfnamefont {J.}~\bibnamefont
  {Kune\ifmmode~\check{s}\else \v{s}\fi{}}},\ }\href {\doibase
  10.1103/PhysRevB.96.045111} {\bibfield  {journal} {\bibinfo  {journal} {Phys.
  Rev. B}\ }\textbf {\bibinfo {volume} {96}},\ \bibinfo {pages} {045111}
  (\bibinfo {year} {2017})}\BibitemShut {NoStop}%
\bibitem [{\citenamefont {Taguchi}\ and\ \citenamefont
  {Panaccione}(2016)}]{Taguchi16book}%
  \BibitemOpen
  \bibfield  {author} {\bibinfo {author} {\bibfnamefont {M.}~\bibnamefont
  {Taguchi}}\ and\ \bibinfo {author} {\bibfnamefont {G.}~\bibnamefont
  {Panaccione}},\ }\enquote {\bibinfo {title} {Depth-dependence of electron
  screening, charge carriers and correlation: Theory and experiments},}\ in\
  \href {\doibase 10.1007/978-3-319-24043-5_9} {\emph {\bibinfo {booktitle}
  {Hard X-ray Photoelectron Spectroscopy (HAXPES)}}},\ \bibinfo {editor}
  {edited by\ \bibinfo {editor} {\bibfnamefont {J.~C.}\ \bibnamefont
  {Woicik}}}\ (\bibinfo  {publisher} {Springer International Publishing},\
  \bibinfo {address} {Cham},\ \bibinfo {year} {2016})\ pp.\ \bibinfo {pages}
  {197--216}\BibitemShut {NoStop}%
\bibitem [{\citenamefont {Taguchi}\ \emph {et~al.}(2008)\citenamefont
  {Taguchi}, \citenamefont {Matsunami}, \citenamefont {Ishida}, \citenamefont
  {Eguchi}, \citenamefont {Chainani}, \citenamefont {Takata}, \citenamefont
  {Yabashi}, \citenamefont {Tamasaku}, \citenamefont {Nishino}, \citenamefont
  {Ishikawa}, \citenamefont {Senba}, \citenamefont {Ohashi},\ and\
  \citenamefont {Shin}}]{taguchi08}%
  \BibitemOpen
  \bibfield  {author} {\bibinfo {author} {\bibfnamefont {M.}~\bibnamefont
  {Taguchi}}, \bibinfo {author} {\bibfnamefont {M.}~\bibnamefont {Matsunami}},
  \bibinfo {author} {\bibfnamefont {Y.}~\bibnamefont {Ishida}}, \bibinfo
  {author} {\bibfnamefont {R.}~\bibnamefont {Eguchi}}, \bibinfo {author}
  {\bibfnamefont {A.}~\bibnamefont {Chainani}}, \bibinfo {author}
  {\bibfnamefont {Y.}~\bibnamefont {Takata}}, \bibinfo {author} {\bibfnamefont
  {M.}~\bibnamefont {Yabashi}}, \bibinfo {author} {\bibfnamefont
  {K.}~\bibnamefont {Tamasaku}}, \bibinfo {author} {\bibfnamefont
  {Y.}~\bibnamefont {Nishino}}, \bibinfo {author} {\bibfnamefont
  {T.}~\bibnamefont {Ishikawa}}, \bibinfo {author} {\bibfnamefont
  {Y.}~\bibnamefont {Senba}}, \bibinfo {author} {\bibfnamefont
  {H.}~\bibnamefont {Ohashi}}, \ and\ \bibinfo {author} {\bibfnamefont
  {S.}~\bibnamefont {Shin}},\ }\href {\doibase 10.1103/PhysRevLett.100.206401}
  {\bibfield  {journal} {\bibinfo  {journal} {Phys. Rev. Lett.}\ }\textbf
  {\bibinfo {volume} {100}},\ \bibinfo {pages} {206401} (\bibinfo {year}
  {2008})}\BibitemShut {NoStop}%
\bibitem [{\citenamefont {Taguchi}\ \emph
  {et~al.}(2005{\natexlab{a}})\citenamefont {Taguchi}, \citenamefont
  {Chainani}, \citenamefont {Horiba}, \citenamefont {Takata}, \citenamefont
  {Yabashi}, \citenamefont {Tamasaku}, \citenamefont {Nishino}, \citenamefont
  {Miwa}, \citenamefont {Ishikawa}, \citenamefont {Takeuchi}, \citenamefont
  {Yamamoto}, \citenamefont {Matsunami}, \citenamefont {Shin}, \citenamefont
  {Yokoya}, \citenamefont {Ikenaga}, \citenamefont {Kobayashi}, \citenamefont
  {Mochiku}, \citenamefont {Hirata}, \citenamefont {Hori}, \citenamefont
  {Ishii}, \citenamefont {Nakamura},\ and\ \citenamefont
  {Suzuki}}]{Taguchi05b}%
  \BibitemOpen
  \bibfield  {author} {\bibinfo {author} {\bibfnamefont {M.}~\bibnamefont
  {Taguchi}}, \bibinfo {author} {\bibfnamefont {A.}~\bibnamefont {Chainani}},
  \bibinfo {author} {\bibfnamefont {K.}~\bibnamefont {Horiba}}, \bibinfo
  {author} {\bibfnamefont {Y.}~\bibnamefont {Takata}}, \bibinfo {author}
  {\bibfnamefont {M.}~\bibnamefont {Yabashi}}, \bibinfo {author} {\bibfnamefont
  {K.}~\bibnamefont {Tamasaku}}, \bibinfo {author} {\bibfnamefont
  {Y.}~\bibnamefont {Nishino}}, \bibinfo {author} {\bibfnamefont
  {D.}~\bibnamefont {Miwa}}, \bibinfo {author} {\bibfnamefont {T.}~\bibnamefont
  {Ishikawa}}, \bibinfo {author} {\bibfnamefont {T.}~\bibnamefont {Takeuchi}},
  \bibinfo {author} {\bibfnamefont {K.}~\bibnamefont {Yamamoto}}, \bibinfo
  {author} {\bibfnamefont {M.}~\bibnamefont {Matsunami}}, \bibinfo {author}
  {\bibfnamefont {S.}~\bibnamefont {Shin}}, \bibinfo {author} {\bibfnamefont
  {T.}~\bibnamefont {Yokoya}}, \bibinfo {author} {\bibfnamefont
  {E.}~\bibnamefont {Ikenaga}}, \bibinfo {author} {\bibfnamefont
  {K.}~\bibnamefont {Kobayashi}}, \bibinfo {author} {\bibfnamefont
  {T.}~\bibnamefont {Mochiku}}, \bibinfo {author} {\bibfnamefont
  {K.}~\bibnamefont {Hirata}}, \bibinfo {author} {\bibfnamefont
  {J.}~\bibnamefont {Hori}}, \bibinfo {author} {\bibfnamefont {K.}~\bibnamefont
  {Ishii}}, \bibinfo {author} {\bibfnamefont {F.}~\bibnamefont {Nakamura}}, \
  and\ \bibinfo {author} {\bibfnamefont {T.}~\bibnamefont {Suzuki}},\ }\href
  {\doibase 10.1103/PhysRevLett.95.177002} {\bibfield  {journal} {\bibinfo
  {journal} {Phys. Rev. Lett.}\ }\textbf {\bibinfo {volume} {95}},\ \bibinfo
  {pages} {177002} (\bibinfo {year} {2005}{\natexlab{a}})}\BibitemShut
  {NoStop}%
\bibitem [{\citenamefont {Horio}\ \emph {et~al.}(2018)\citenamefont {Horio},
  \citenamefont {Krockenberger}, \citenamefont {Yamamoto}, \citenamefont
  {Yokoyama}, \citenamefont {Takubo}, \citenamefont {Hirata}, \citenamefont
  {Sakamoto}, \citenamefont {Koshiishi}, \citenamefont {Yasui}, \citenamefont
  {Ikenaga}, \citenamefont {Shin}, \citenamefont {Yamamoto}, \citenamefont
  {Wadati},\ and\ \citenamefont {Fujimori}}]{Horio18}%
  \BibitemOpen
  \bibfield  {author} {\bibinfo {author} {\bibfnamefont {M.}~\bibnamefont
  {Horio}}, \bibinfo {author} {\bibfnamefont {Y.}~\bibnamefont
  {Krockenberger}}, \bibinfo {author} {\bibfnamefont {K.}~\bibnamefont
  {Yamamoto}}, \bibinfo {author} {\bibfnamefont {Y.}~\bibnamefont {Yokoyama}},
  \bibinfo {author} {\bibfnamefont {K.}~\bibnamefont {Takubo}}, \bibinfo
  {author} {\bibfnamefont {Y.}~\bibnamefont {Hirata}}, \bibinfo {author}
  {\bibfnamefont {S.}~\bibnamefont {Sakamoto}}, \bibinfo {author}
  {\bibfnamefont {K.}~\bibnamefont {Koshiishi}}, \bibinfo {author}
  {\bibfnamefont {A.}~\bibnamefont {Yasui}}, \bibinfo {author} {\bibfnamefont
  {E.}~\bibnamefont {Ikenaga}}, \bibinfo {author} {\bibfnamefont
  {S.}~\bibnamefont {Shin}}, \bibinfo {author} {\bibfnamefont {H.}~\bibnamefont
  {Yamamoto}}, \bibinfo {author} {\bibfnamefont {H.}~\bibnamefont {Wadati}}, \
  and\ \bibinfo {author} {\bibfnamefont {A.}~\bibnamefont {Fujimori}},\ }\href
  {\doibase 10.1103/PhysRevLett.120.257001} {\bibfield  {journal} {\bibinfo
  {journal} {Phys. Rev. Lett.}\ }\textbf {\bibinfo {volume} {120}},\ \bibinfo
  {pages} {257001} (\bibinfo {year} {2018})}\BibitemShut {NoStop}%
\bibitem [{\citenamefont {Okada}\ and\ \citenamefont {Kotani}(1995)}]{Okada95}%
  \BibitemOpen
  \bibfield  {author} {\bibinfo {author} {\bibfnamefont {K.}~\bibnamefont
  {Okada}}\ and\ \bibinfo {author} {\bibfnamefont {A.}~\bibnamefont {Kotani}},\
  }\href {\doibase 10.1103/PhysRevB.52.4794} {\bibfield  {journal} {\bibinfo
  {journal} {Phys. Rev. B}\ }\textbf {\bibinfo {volume} {52}},\ \bibinfo
  {pages} {4794} (\bibinfo {year} {1995})}\BibitemShut {NoStop}%
\bibitem [{\citenamefont {van Veenendaal}\ \emph {et~al.}(1994)\citenamefont
  {van Veenendaal}, \citenamefont {Sawatzky},\ and\ \citenamefont
  {Groen}}]{Veenendaal94}%
  \BibitemOpen
  \bibfield  {author} {\bibinfo {author} {\bibfnamefont {M.~A.}\ \bibnamefont
  {van Veenendaal}}, \bibinfo {author} {\bibfnamefont {G.~A.}\ \bibnamefont
  {Sawatzky}}, \ and\ \bibinfo {author} {\bibfnamefont {W.~A.}\ \bibnamefont
  {Groen}},\ }\href {\doibase 10.1103/PhysRevB.49.1407} {\bibfield  {journal}
  {\bibinfo  {journal} {Phys. Rev. B}\ }\textbf {\bibinfo {volume} {49}},\
  \bibinfo {pages} {1407} (\bibinfo {year} {1994})}\BibitemShut {NoStop}%
\bibitem [{\citenamefont {Taguchi}\ \emph
  {et~al.}(2005{\natexlab{b}})\citenamefont {Taguchi}, \citenamefont
  {Chainani}, \citenamefont {Kamakura}, \citenamefont {Horiba}, \citenamefont
  {Takata}, \citenamefont {Yabashi}, \citenamefont {Tamasaku}, \citenamefont
  {Nishino}, \citenamefont {Miwa}, \citenamefont {Ishikawa}, \citenamefont
  {Shin}, \citenamefont {Ikenaga}, \citenamefont {Yokoya}, \citenamefont
  {Kobayashi}, \citenamefont {Mochiku}, \citenamefont {Hirata},\ and\
  \citenamefont {Motoya}}]{taguchi05}%
  \BibitemOpen
  \bibfield  {author} {\bibinfo {author} {\bibfnamefont {M.}~\bibnamefont
  {Taguchi}}, \bibinfo {author} {\bibfnamefont {A.}~\bibnamefont {Chainani}},
  \bibinfo {author} {\bibfnamefont {N.}~\bibnamefont {Kamakura}}, \bibinfo
  {author} {\bibfnamefont {K.}~\bibnamefont {Horiba}}, \bibinfo {author}
  {\bibfnamefont {Y.}~\bibnamefont {Takata}}, \bibinfo {author} {\bibfnamefont
  {M.}~\bibnamefont {Yabashi}}, \bibinfo {author} {\bibfnamefont
  {K.}~\bibnamefont {Tamasaku}}, \bibinfo {author} {\bibfnamefont
  {Y.}~\bibnamefont {Nishino}}, \bibinfo {author} {\bibfnamefont
  {D.}~\bibnamefont {Miwa}}, \bibinfo {author} {\bibfnamefont {T.}~\bibnamefont
  {Ishikawa}}, \bibinfo {author} {\bibfnamefont {S.}~\bibnamefont {Shin}},
  \bibinfo {author} {\bibfnamefont {E.}~\bibnamefont {Ikenaga}}, \bibinfo
  {author} {\bibfnamefont {T.}~\bibnamefont {Yokoya}}, \bibinfo {author}
  {\bibfnamefont {K.}~\bibnamefont {Kobayashi}}, \bibinfo {author}
  {\bibfnamefont {T.}~\bibnamefont {Mochiku}}, \bibinfo {author} {\bibfnamefont
  {K.}~\bibnamefont {Hirata}}, \ and\ \bibinfo {author} {\bibfnamefont
  {K.}~\bibnamefont {Motoya}},\ }\href {\doibase 10.1103/PhysRevB.71.155102}
  {\bibfield  {journal} {\bibinfo  {journal} {Phys. Rev. B}\ }\textbf {\bibinfo
  {volume} {71}},\ \bibinfo {pages} {155102} (\bibinfo {year}
  {2005}{\natexlab{b}})}\BibitemShut {NoStop}%
\bibitem [{\citenamefont {Metzner}\ and\ \citenamefont
  {Vollhardt}(1989)}]{metzner89}%
  \BibitemOpen
  \bibfield  {author} {\bibinfo {author} {\bibfnamefont {W.}~\bibnamefont
  {Metzner}}\ and\ \bibinfo {author} {\bibfnamefont {D.}~\bibnamefont
  {Vollhardt}},\ }\href {\doibase 10.1103/PhysRevLett.62.324} {\bibfield
  {journal} {\bibinfo  {journal} {Phys. Rev. Lett.}\ }\textbf {\bibinfo
  {volume} {62}},\ \bibinfo {pages} {324} (\bibinfo {year} {1989})}\BibitemShut
  {NoStop}%
\bibitem [{\citenamefont {Georges}\ \emph {et~al.}(1996)\citenamefont
  {Georges}, \citenamefont {Kotliar}, \citenamefont {Krauth},\ and\
  \citenamefont {Rozenberg}}]{georges96}%
  \BibitemOpen
  \bibfield  {author} {\bibinfo {author} {\bibfnamefont {A.}~\bibnamefont
  {Georges}}, \bibinfo {author} {\bibfnamefont {G.}~\bibnamefont {Kotliar}},
  \bibinfo {author} {\bibfnamefont {W.}~\bibnamefont {Krauth}}, \ and\ \bibinfo
  {author} {\bibfnamefont {M.~J.}\ \bibnamefont {Rozenberg}},\ }\href {\doibase
  10.1103/RevModPhys.68.13} {\bibfield  {journal} {\bibinfo  {journal} {Rev.
  Mod. Phys.}\ }\textbf {\bibinfo {volume} {68}},\ \bibinfo {pages} {13}
  (\bibinfo {year} {1996})}\BibitemShut {NoStop}%
\bibitem [{\citenamefont {Kotliar}\ \emph {et~al.}(2006)\citenamefont
  {Kotliar}, \citenamefont {Savrasov}, \citenamefont {Haule}, \citenamefont
  {Oudovenko}, \citenamefont {Parcollet},\ and\ \citenamefont
  {Marianetti}}]{kotliar06}%
  \BibitemOpen
  \bibfield  {author} {\bibinfo {author} {\bibfnamefont {G.}~\bibnamefont
  {Kotliar}}, \bibinfo {author} {\bibfnamefont {S.~Y.}\ \bibnamefont
  {Savrasov}}, \bibinfo {author} {\bibfnamefont {K.}~\bibnamefont {Haule}},
  \bibinfo {author} {\bibfnamefont {V.~S.}\ \bibnamefont {Oudovenko}}, \bibinfo
  {author} {\bibfnamefont {O.}~\bibnamefont {Parcollet}}, \ and\ \bibinfo
  {author} {\bibfnamefont {C.~A.}\ \bibnamefont {Marianetti}},\ }\href
  {\doibase 10.1103/RevModPhys.78.865} {\bibfield  {journal} {\bibinfo
  {journal} {Rev. Mod. Phys.}\ }\textbf {\bibinfo {volume} {78}},\ \bibinfo
  {pages} {865} (\bibinfo {year} {2006})}\BibitemShut {NoStop}%
\bibitem [{\citenamefont {Hariki}\ \emph {et~al.}(2018)\citenamefont {Hariki},
  \citenamefont {Winder},\ and\ \citenamefont {Kune\ifmmode~\check{s}\else
  \v{s}\fi{}}}]{Hariki18}%
  \BibitemOpen
  \bibfield  {author} {\bibinfo {author} {\bibfnamefont {A.}~\bibnamefont
  {Hariki}}, \bibinfo {author} {\bibfnamefont {M.}~\bibnamefont {Winder}}, \
  and\ \bibinfo {author} {\bibfnamefont {J.}~\bibnamefont
  {Kune\ifmmode~\check{s}\else \v{s}\fi{}}},\ }\href {\doibase
  10.1103/PhysRevLett.121.126403} {\bibfield  {journal} {\bibinfo  {journal}
  {Phys. Rev. Lett.}\ }\textbf {\bibinfo {volume} {121}},\ \bibinfo {pages}
  {126403} (\bibinfo {year} {2018})}\BibitemShut {NoStop}%
\bibitem [{\citenamefont {Hariki}\ \emph {et~al.}(2020)\citenamefont {Hariki},
  \citenamefont {Winder}, \citenamefont {Uozumi},\ and\ \citenamefont
  {Kune\ifmmode~\check{s}\else \v{s}\fi{}}}]{Hariki20}%
  \BibitemOpen
  \bibfield  {author} {\bibinfo {author} {\bibfnamefont {A.}~\bibnamefont
  {Hariki}}, \bibinfo {author} {\bibfnamefont {M.}~\bibnamefont {Winder}},
  \bibinfo {author} {\bibfnamefont {T.}~\bibnamefont {Uozumi}}, \ and\ \bibinfo
  {author} {\bibfnamefont {J.}~\bibnamefont {Kune\ifmmode~\check{s}\else
  \v{s}\fi{}}},\ }\href {\doibase 10.1103/PhysRevB.101.115130} {\bibfield
  {journal} {\bibinfo  {journal} {Phys. Rev. B}\ }\textbf {\bibinfo {volume}
  {101}},\ \bibinfo {pages} {115130} (\bibinfo {year} {2020})}\BibitemShut
  {NoStop}%
\bibitem [{\citenamefont {Ghiasi}\ \emph {et~al.}(2019)\citenamefont {Ghiasi},
  \citenamefont {Hariki}, \citenamefont {Winder}, \citenamefont
  {Kune\ifmmode~\check{s}\else \v{s}\fi{}}, \citenamefont {Regoutz},
  \citenamefont {Lee}, \citenamefont {Hu}, \citenamefont {Rueff},\ and\
  \citenamefont {de~Groot}}]{Ghiasi19}%
  \BibitemOpen
  \bibfield  {author} {\bibinfo {author} {\bibfnamefont {M.}~\bibnamefont
  {Ghiasi}}, \bibinfo {author} {\bibfnamefont {A.}~\bibnamefont {Hariki}},
  \bibinfo {author} {\bibfnamefont {M.}~\bibnamefont {Winder}}, \bibinfo
  {author} {\bibfnamefont {J.}~\bibnamefont {Kune\ifmmode~\check{s}\else
  \v{s}\fi{}}}, \bibinfo {author} {\bibfnamefont {A.}~\bibnamefont {Regoutz}},
  \bibinfo {author} {\bibfnamefont {T.-L.}\ \bibnamefont {Lee}}, \bibinfo
  {author} {\bibfnamefont {Y.}~\bibnamefont {Hu}}, \bibinfo {author}
  {\bibfnamefont {J.-P.}\ \bibnamefont {Rueff}}, \ and\ \bibinfo {author}
  {\bibfnamefont {F.~M.~F.}\ \bibnamefont {de~Groot}},\ }\href {\doibase
  10.1103/PhysRevB.100.075146} {\bibfield  {journal} {\bibinfo  {journal}
  {Phys. Rev. B}\ }\textbf {\bibinfo {volume} {100}},\ \bibinfo {pages}
  {075146} (\bibinfo {year} {2019})}\BibitemShut {NoStop}%
\bibitem [{\citenamefont {Koloren\v{c}}(2018)}]{Jindrich18}%
  \BibitemOpen
  \bibfield  {author} {\bibinfo {author} {\bibfnamefont {J.}~\bibnamefont
  {Koloren\v{c}}},\ }\href {\doibase
  https://doi.org/10.1016/j.physb.2017.08.069} {\bibfield  {journal} {\bibinfo
  {journal} {Physica B Condens. Matter.}\ }\textbf {\bibinfo {volume} {536}},\
  \bibinfo {pages} {695 } (\bibinfo {year} {2018})}\BibitemShut {NoStop}%
\bibitem [{\citenamefont {Wang}\ \emph {et~al.}(2020)\citenamefont {Wang},
  \citenamefont {Kang}, \citenamefont {Miao},\ and\ \citenamefont
  {Kotliar}}]{Wang20}%
  \BibitemOpen
  \bibfield  {author} {\bibinfo {author} {\bibfnamefont {Y.}~\bibnamefont
  {Wang}}, \bibinfo {author} {\bibfnamefont {C.-J.}\ \bibnamefont {Kang}},
  \bibinfo {author} {\bibfnamefont {H.}~\bibnamefont {Miao}}, \ and\ \bibinfo
  {author} {\bibfnamefont {G.}~\bibnamefont {Kotliar}},\ }\href {\doibase
  10.1103/PhysRevB.102.161118} {\bibfield  {journal} {\bibinfo  {journal}
  {Phys. Rev. B}\ }\textbf {\bibinfo {volume} {102}},\ \bibinfo {pages}
  {161118} (\bibinfo {year} {2020})}\BibitemShut {NoStop}%
\bibitem [{\citenamefont {Kang}\ and\ \citenamefont {Kotliar}(2021)}]{Kang20}%
  \BibitemOpen
  \bibfield  {author} {\bibinfo {author} {\bibfnamefont {C.-J.}\ \bibnamefont
  {Kang}}\ and\ \bibinfo {author} {\bibfnamefont {G.}~\bibnamefont {Kotliar}},\
  }\href {\doibase 10.1103/PhysRevLett.126.127401} {\bibfield  {journal}
  {\bibinfo  {journal} {Phys. Rev. Lett.}\ }\textbf {\bibinfo {volume} {126}},\
  \bibinfo {pages} {127401} (\bibinfo {year} {2021})}\BibitemShut {NoStop}%
\bibitem [{\citenamefont {Petocchi}\ \emph {et~al.}(2020)\citenamefont
  {Petocchi}, \citenamefont {Christiansson}, \citenamefont {Nilsson},
  \citenamefont {Aryasetiawan},\ and\ \citenamefont {Werner}}]{Petocchi20}%
  \BibitemOpen
  \bibfield  {author} {\bibinfo {author} {\bibfnamefont {F.}~\bibnamefont
  {Petocchi}}, \bibinfo {author} {\bibfnamefont {V.}~\bibnamefont
  {Christiansson}}, \bibinfo {author} {\bibfnamefont {F.}~\bibnamefont
  {Nilsson}}, \bibinfo {author} {\bibfnamefont {F.}~\bibnamefont
  {Aryasetiawan}}, \ and\ \bibinfo {author} {\bibfnamefont {P.}~\bibnamefont
  {Werner}},\ }\href {\doibase 10.1103/PhysRevX.10.041047} {\bibfield
  {journal} {\bibinfo  {journal} {Phys. Rev. X}\ }\textbf {\bibinfo {volume}
  {10}},\ \bibinfo {pages} {041047} (\bibinfo {year} {2020})}\BibitemShut
  {NoStop}%
\bibitem [{\citenamefont {Lechermann}(2020)}]{Lechermann20b}%
  \BibitemOpen
  \bibfield  {author} {\bibinfo {author} {\bibfnamefont {F.}~\bibnamefont
  {Lechermann}},\ }\href {\doibase 10.1103/PhysRevX.10.041002} {\bibfield
  {journal} {\bibinfo  {journal} {Phys. Rev. X}\ }\textbf {\bibinfo {volume}
  {10}},\ \bibinfo {pages} {041002} (\bibinfo {year} {2020})}\BibitemShut
  {NoStop}%
\bibitem [{\citenamefont {Karp}\ \emph
  {et~al.}(2020{\natexlab{a}})\citenamefont {Karp}, \citenamefont {Botana},
  \citenamefont {Norman}, \citenamefont {Park}, \citenamefont {Zingl},\ and\
  \citenamefont {Millis}}]{Karp20}%
  \BibitemOpen
  \bibfield  {author} {\bibinfo {author} {\bibfnamefont {J.}~\bibnamefont
  {Karp}}, \bibinfo {author} {\bibfnamefont {A.~S.}\ \bibnamefont {Botana}},
  \bibinfo {author} {\bibfnamefont {M.~R.}\ \bibnamefont {Norman}}, \bibinfo
  {author} {\bibfnamefont {H.}~\bibnamefont {Park}}, \bibinfo {author}
  {\bibfnamefont {M.}~\bibnamefont {Zingl}}, \ and\ \bibinfo {author}
  {\bibfnamefont {A.}~\bibnamefont {Millis}},\ }\href {\doibase
  10.1103/PhysRevX.10.021061} {\bibfield  {journal} {\bibinfo  {journal} {Phys.
  Rev. X}\ }\textbf {\bibinfo {volume} {10}},\ \bibinfo {pages} {021061}
  (\bibinfo {year} {2020}{\natexlab{a}})}\BibitemShut {NoStop}%
\bibitem [{\citenamefont {Kitatani}\ \emph {et~al.}(2020)\citenamefont
  {Kitatani}, \citenamefont {Si}, \citenamefont {Janson}, \citenamefont
  {Arita}, \citenamefont {Zhong},\ and\ \citenamefont {Held}}]{Kitatani20}%
  \BibitemOpen
  \bibfield  {author} {\bibinfo {author} {\bibfnamefont {M.}~\bibnamefont
  {Kitatani}}, \bibinfo {author} {\bibfnamefont {L.}~\bibnamefont {Si}},
  \bibinfo {author} {\bibfnamefont {O.}~\bibnamefont {Janson}}, \bibinfo
  {author} {\bibfnamefont {R.}~\bibnamefont {Arita}}, \bibinfo {author}
  {\bibfnamefont {Z.}~\bibnamefont {Zhong}}, \ and\ \bibinfo {author}
  {\bibfnamefont {K.}~\bibnamefont {Held}},\ }\href {\doibase
  10.1038/s41535-020-00260-y} {\bibfield  {journal} {\bibinfo  {journal} {npj
  Quantum Materials}\ }\textbf {\bibinfo {volume} {5}},\ \bibinfo {pages} {59}
  (\bibinfo {year} {2020})}\BibitemShut {NoStop}%
\bibitem [{\citenamefont {Karp}\ \emph
  {et~al.}(2020{\natexlab{b}})\citenamefont {Karp}, \citenamefont {Hampel},
  \citenamefont {Zingl}, \citenamefont {Botana}, \citenamefont {Park},
  \citenamefont {Norman},\ and\ \citenamefont {Millis}}]{Karp20b}%
  \BibitemOpen
  \bibfield  {author} {\bibinfo {author} {\bibfnamefont {J.}~\bibnamefont
  {Karp}}, \bibinfo {author} {\bibfnamefont {A.}~\bibnamefont {Hampel}},
  \bibinfo {author} {\bibfnamefont {M.}~\bibnamefont {Zingl}}, \bibinfo
  {author} {\bibfnamefont {A.~S.}\ \bibnamefont {Botana}}, \bibinfo {author}
  {\bibfnamefont {H.}~\bibnamefont {Park}}, \bibinfo {author} {\bibfnamefont
  {M.~R.}\ \bibnamefont {Norman}}, \ and\ \bibinfo {author} {\bibfnamefont
  {A.~J.}\ \bibnamefont {Millis}},\ }\href {\doibase
  10.1103/PhysRevB.102.245130} {\bibfield  {journal} {\bibinfo  {journal}
  {Phys. Rev. B}\ }\textbf {\bibinfo {volume} {102}},\ \bibinfo {pages}
  {245130} (\bibinfo {year} {2020}{\natexlab{b}})}\BibitemShut {NoStop}%
\bibitem [{\citenamefont {{Karp}}\ \emph {et~al.}(2021)\citenamefont {{Karp}},
  \citenamefont {{Hampel}},\ and\ \citenamefont {{Millis}}}]{Karp21}%
  \BibitemOpen
  \bibfield  {author} {\bibinfo {author} {\bibfnamefont {J.}~\bibnamefont
  {{Karp}}}, \bibinfo {author} {\bibfnamefont {A.}~\bibnamefont {{Hampel}}}, \
  and\ \bibinfo {author} {\bibfnamefont {A.~J.}\ \bibnamefont {{Millis}}},\
  }\href@noop {} {\  (\bibinfo {year} {2021})},\ \Eprint
  {http://arxiv.org/abs/2102.08522} {arXiv:2102.08522} \BibitemShut {NoStop}%
\bibitem [{\citenamefont {Kune{\v{s}}}\ \emph {et~al.}(2009)\citenamefont
  {Kune{\v{s}}}, \citenamefont {Leonov}, \citenamefont {Kollar}, \citenamefont
  {Byczuk}, \citenamefont {Anisimov},\ and\ \citenamefont
  {Vollhardt}}]{kunes09}%
  \BibitemOpen
  \bibfield  {author} {\bibinfo {author} {\bibfnamefont {J.}~\bibnamefont
  {Kune{\v{s}}}}, \bibinfo {author} {\bibfnamefont {I.}~\bibnamefont {Leonov}},
  \bibinfo {author} {\bibfnamefont {M.}~\bibnamefont {Kollar}}, \bibinfo
  {author} {\bibfnamefont {K.}~\bibnamefont {Byczuk}}, \bibinfo {author}
  {\bibfnamefont {V.~I.}\ \bibnamefont {Anisimov}}, \ and\ \bibinfo {author}
  {\bibfnamefont {D.}~\bibnamefont {Vollhardt}},\ }\href {\doibase
  10.1140/epjst/e2010-01209-0} {\bibfield  {journal} {\bibinfo  {journal} {Eur.
  Phys. J. Spec. Top.}\ }\textbf {\bibinfo {volume} {180}},\ \bibinfo {pages}
  {5} (\bibinfo {year} {2009})}\BibitemShut {NoStop}%
\bibitem [{\citenamefont {Hayward}\ \emph {et~al.}(1999)\citenamefont
  {Hayward}, \citenamefont {Green}, \citenamefont {Rosseinsky},\ and\
  \citenamefont {Sloan}}]{Hayward99}%
  \BibitemOpen
  \bibfield  {author} {\bibinfo {author} {\bibfnamefont {M.~A.}\ \bibnamefont
  {Hayward}}, \bibinfo {author} {\bibfnamefont {M.~A.}\ \bibnamefont {Green}},
  \bibinfo {author} {\bibfnamefont {M.~J.}\ \bibnamefont {Rosseinsky}}, \ and\
  \bibinfo {author} {\bibfnamefont {J.}~\bibnamefont {Sloan}},\ }\href
  {\doibase 10.1021/ja991573i} {\bibfield  {journal} {\bibinfo  {journal}
  {Journal of the American Chemical Society}\ }\textbf {\bibinfo {volume}
  {121}},\ \bibinfo {pages} {8843} (\bibinfo {year} {1999})},\ \Eprint
  {http://arxiv.org/abs/https://doi.org/10.1021/ja991573i}
  {https://doi.org/10.1021/ja991573i} \BibitemShut {NoStop}%
\bibitem [{\citenamefont {Blaha}\ \emph {et~al.}()\citenamefont {Blaha},
  \citenamefont {Schwarz}, \citenamefont {Madsen}, \citenamefont {Kvasnicka},\
  and\ \citenamefont {Luitz}}]{wien2k}%
  \BibitemOpen
  \bibfield  {author} {\bibinfo {author} {\bibfnamefont {P.}~\bibnamefont
  {Blaha}}, \bibinfo {author} {\bibfnamefont {K.}~\bibnamefont {Schwarz}},
  \bibinfo {author} {\bibfnamefont {G.}~\bibnamefont {Madsen}}, \bibinfo
  {author} {\bibfnamefont {D.}~\bibnamefont {Kvasnicka}}, \ and\ \bibinfo
  {author} {\bibfnamefont {J.}~\bibnamefont {Luitz}},\ }\href@noop {} {\emph
  {\bibinfo {title} {WIEN2k, An Augmented Plane Wave + Local Orbitals Program
  for Calculating Crystal Properties}}}\ (\bibinfo  {publisher} {Karlheinz
  Schwarz, Techn. Universitat Wien, Austria, 2001, ISBN
  3-9501031-1-2})\BibitemShut {NoStop}%
\bibitem [{Note1()}]{Note1}%
  \BibitemOpen
  \bibinfo {note} {The Nd 4$f$ states in NdNiO$_2$ are treated as
  partially-filled core states.}\BibitemShut {Stop}%
\bibitem [{\citenamefont {Kune\v{s}}\ \emph {et~al.}(2010)\citenamefont
  {Kune\v{s}}, \citenamefont {Arita}, \citenamefont {Wissgott}, \citenamefont
  {Toschi}, \citenamefont {Ikeda},\ and\ \citenamefont {Held}}]{wien2wannier}%
  \BibitemOpen
  \bibfield  {author} {\bibinfo {author} {\bibfnamefont {J.}~\bibnamefont
  {Kune\v{s}}}, \bibinfo {author} {\bibfnamefont {R.}~\bibnamefont {Arita}},
  \bibinfo {author} {\bibfnamefont {P.}~\bibnamefont {Wissgott}}, \bibinfo
  {author} {\bibfnamefont {A.}~\bibnamefont {Toschi}}, \bibinfo {author}
  {\bibfnamefont {H.}~\bibnamefont {Ikeda}}, \ and\ \bibinfo {author}
  {\bibfnamefont {K.}~\bibnamefont {Held}},\ }\href {\doibase
  http://dx.doi.org/10.1016/j.cpc.2010.08.005} {\bibfield  {journal} {\bibinfo
  {journal} {Comput. Phys. Commun.}\ }\textbf {\bibinfo {volume} {181}},\
  \bibinfo {pages} {1888 } (\bibinfo {year} {2010})}\BibitemShut {NoStop}%
\bibitem [{\citenamefont {Mostofi}\ \emph {et~al.}(2014)\citenamefont
  {Mostofi}, \citenamefont {Yates}, \citenamefont {Pizzi}, \citenamefont {Lee},
  \citenamefont {Souza}, \citenamefont {Vanderbilt},\ and\ \citenamefont
  {Marzari}}]{wannier90}%
  \BibitemOpen
  \bibfield  {author} {\bibinfo {author} {\bibfnamefont {A.~A.}\ \bibnamefont
  {Mostofi}}, \bibinfo {author} {\bibfnamefont {J.~R.}\ \bibnamefont {Yates}},
  \bibinfo {author} {\bibfnamefont {G.}~\bibnamefont {Pizzi}}, \bibinfo
  {author} {\bibfnamefont {Y.-S.}\ \bibnamefont {Lee}}, \bibinfo {author}
  {\bibfnamefont {I.}~\bibnamefont {Souza}}, \bibinfo {author} {\bibfnamefont
  {D.}~\bibnamefont {Vanderbilt}}, \ and\ \bibinfo {author} {\bibfnamefont
  {N.}~\bibnamefont {Marzari}},\ }\href {\doibase
  http://dx.doi.org/10.1016/j.cpc.2014.05.003} {\bibfield  {journal} {\bibinfo
  {journal} {Comput. Phys. Commun.}\ }\textbf {\bibinfo {volume} {185}},\
  \bibinfo {pages} {2309 } (\bibinfo {year} {2014})}\BibitemShut {NoStop}%
\bibitem [{\citenamefont {Ryee}\ \emph {et~al.}(2020)\citenamefont {Ryee},
  \citenamefont {Yoon}, \citenamefont {Kim}, \citenamefont {Jeong},\ and\
  \citenamefont {Han}}]{Ryee20}%
  \BibitemOpen
  \bibfield  {author} {\bibinfo {author} {\bibfnamefont {S.}~\bibnamefont
  {Ryee}}, \bibinfo {author} {\bibfnamefont {H.}~\bibnamefont {Yoon}}, \bibinfo
  {author} {\bibfnamefont {T.~J.}\ \bibnamefont {Kim}}, \bibinfo {author}
  {\bibfnamefont {M.~Y.}\ \bibnamefont {Jeong}}, \ and\ \bibinfo {author}
  {\bibfnamefont {M.~J.}\ \bibnamefont {Han}},\ }\href {\doibase
  10.1103/PhysRevB.101.064513} {\bibfield  {journal} {\bibinfo  {journal}
  {Phys. Rev. B}\ }\textbf {\bibinfo {volume} {101}},\ \bibinfo {pages}
  {064513} (\bibinfo {year} {2020})}\BibitemShut {NoStop}%
\bibitem [{\citenamefont {Werner}\ \emph {et~al.}(2006)\citenamefont {Werner},
  \citenamefont {Comanac}, \citenamefont {de' Medici}, \citenamefont {Troyer},\
  and\ \citenamefont {Millis}}]{werner06}%
  \BibitemOpen
  \bibfield  {author} {\bibinfo {author} {\bibfnamefont {P.}~\bibnamefont
  {Werner}}, \bibinfo {author} {\bibfnamefont {A.}~\bibnamefont {Comanac}},
  \bibinfo {author} {\bibfnamefont {L.}~\bibnamefont {de' Medici}}, \bibinfo
  {author} {\bibfnamefont {M.}~\bibnamefont {Troyer}}, \ and\ \bibinfo {author}
  {\bibfnamefont {A.~J.}\ \bibnamefont {Millis}},\ }\href {\doibase
  10.1103/PhysRevLett.97.076405} {\bibfield  {journal} {\bibinfo  {journal}
  {Phys. Rev. Lett.}\ }\textbf {\bibinfo {volume} {97}},\ \bibinfo {pages}
  {076405} (\bibinfo {year} {2006})}\BibitemShut {NoStop}%
\bibitem [{\citenamefont {Boehnke}\ \emph {et~al.}(2011)\citenamefont
  {Boehnke}, \citenamefont {Hafermann}, \citenamefont {Ferrero}, \citenamefont
  {Lechermann},\ and\ \citenamefont {Parcollet}}]{boehnke11}%
  \BibitemOpen
  \bibfield  {author} {\bibinfo {author} {\bibfnamefont {L.}~\bibnamefont
  {Boehnke}}, \bibinfo {author} {\bibfnamefont {H.}~\bibnamefont {Hafermann}},
  \bibinfo {author} {\bibfnamefont {M.}~\bibnamefont {Ferrero}}, \bibinfo
  {author} {\bibfnamefont {F.}~\bibnamefont {Lechermann}}, \ and\ \bibinfo
  {author} {\bibfnamefont {O.}~\bibnamefont {Parcollet}},\ }\href {\doibase
  10.1103/PhysRevB.84.075145} {\bibfield  {journal} {\bibinfo  {journal} {Phys.
  Rev. B}\ }\textbf {\bibinfo {volume} {84}},\ \bibinfo {pages} {075145}
  (\bibinfo {year} {2011})}\BibitemShut {NoStop}%
\bibitem [{\citenamefont {Hafermann}\ \emph {et~al.}(2012)\citenamefont
  {Hafermann}, \citenamefont {Patton},\ and\ \citenamefont
  {Werner}}]{hafermann12}%
  \BibitemOpen
  \bibfield  {author} {\bibinfo {author} {\bibfnamefont {H.}~\bibnamefont
  {Hafermann}}, \bibinfo {author} {\bibfnamefont {K.~R.}\ \bibnamefont
  {Patton}}, \ and\ \bibinfo {author} {\bibfnamefont {P.}~\bibnamefont
  {Werner}},\ }\href {\doibase 10.1103/PhysRevB.85.205106} {\bibfield
  {journal} {\bibinfo  {journal} {Phys. Rev. B}\ }\textbf {\bibinfo {volume}
  {85}},\ \bibinfo {pages} {205106} (\bibinfo {year} {2012})}\BibitemShut
  {NoStop}%
\bibitem [{\citenamefont {Hariki}\ \emph {et~al.}(2015)\citenamefont {Hariki},
  \citenamefont {Yamanaka},\ and\ \citenamefont {Uozumi}}]{Hariki15}%
  \BibitemOpen
  \bibfield  {author} {\bibinfo {author} {\bibfnamefont {A.}~\bibnamefont
  {Hariki}}, \bibinfo {author} {\bibfnamefont {A.}~\bibnamefont {Yamanaka}}, \
  and\ \bibinfo {author} {\bibfnamefont {T.}~\bibnamefont {Uozumi}},\ }\href
  {\doibase 10.7566/JPSJ.84.073706} {\bibfield  {journal} {\bibinfo  {journal}
  {J. Phys. Soc. Jpn.}\ }\textbf {\bibinfo {volume} {84}},\ \bibinfo {pages}
  {073706} (\bibinfo {year} {2015})}\BibitemShut {NoStop}%
\bibitem [{\citenamefont {Jarrell}\ and\ \citenamefont
  {Gubernatis}(1996)}]{jarrell96}%
  \BibitemOpen
  \bibfield  {author} {\bibinfo {author} {\bibfnamefont {M.}~\bibnamefont
  {Jarrell}}\ and\ \bibinfo {author} {\bibfnamefont {J.}~\bibnamefont
  {Gubernatis}},\ }\href {\doibase
  http://dx.doi.org/10.1016/0370-1573(95)00074-7} {\bibfield  {journal}
  {\bibinfo  {journal} {Phys. Rep.}\ }\textbf {\bibinfo {volume} {269}},\
  \bibinfo {pages} {133 } (\bibinfo {year} {1996})}\BibitemShut {NoStop}%
\bibitem [{\citenamefont {Winder}\ \emph {et~al.}(2020)\citenamefont {Winder},
  \citenamefont {Hariki},\ and\ \citenamefont {Kune\ifmmode~\check{s}\else
  \v{s}\fi{}}}]{Winder20}%
  \BibitemOpen
  \bibfield  {author} {\bibinfo {author} {\bibfnamefont {M.}~\bibnamefont
  {Winder}}, \bibinfo {author} {\bibfnamefont {A.}~\bibnamefont {Hariki}}, \
  and\ \bibinfo {author} {\bibfnamefont {J.}~\bibnamefont
  {Kune\ifmmode~\check{s}\else \v{s}\fi{}}},\ }\href {\doibase
  10.1103/PhysRevB.102.085155} {\bibfield  {journal} {\bibinfo  {journal}
  {Phys. Rev. B}\ }\textbf {\bibinfo {volume} {102}},\ \bibinfo {pages}
  {085155} (\bibinfo {year} {2020})}\BibitemShut {NoStop}%
\bibitem [{\citenamefont {Karolak}\ \emph {et~al.}(2010)\citenamefont
  {Karolak}, \citenamefont {Ulm}, \citenamefont {Wehling}, \citenamefont
  {Mazurenko}, \citenamefont {Poteryaev},\ and\ \citenamefont
  {Lichtenstein}}]{karolak10}%
  \BibitemOpen
  \bibfield  {author} {\bibinfo {author} {\bibfnamefont {M.}~\bibnamefont
  {Karolak}}, \bibinfo {author} {\bibfnamefont {G.}~\bibnamefont {Ulm}},
  \bibinfo {author} {\bibfnamefont {T.}~\bibnamefont {Wehling}}, \bibinfo
  {author} {\bibfnamefont {V.}~\bibnamefont {Mazurenko}}, \bibinfo {author}
  {\bibfnamefont {A.}~\bibnamefont {Poteryaev}}, \ and\ \bibinfo {author}
  {\bibfnamefont {A.}~\bibnamefont {Lichtenstein}},\ }\href {\doibase
  https://doi.org/10.1016/j.elspec.2010.05.021} {\bibfield  {journal} {\bibinfo
   {journal} {J. Electron. Spectrosc. Relat. Phenom.}\ }\textbf {\bibinfo
  {volume} {181}},\ \bibinfo {pages} {11 } (\bibinfo {year}
  {2010})}\BibitemShut {NoStop}%
\bibitem [{\citenamefont {Haule}(2015)}]{Haule15}%
  \BibitemOpen
  \bibfield  {author} {\bibinfo {author} {\bibfnamefont {K.}~\bibnamefont
  {Haule}},\ }\href {\doibase 10.1103/PhysRevLett.115.196403} {\bibfield
  {journal} {\bibinfo  {journal} {Phys. Rev. Lett.}\ }\textbf {\bibinfo
  {volume} {115}},\ \bibinfo {pages} {196403} (\bibinfo {year}
  {2015})}\BibitemShut {NoStop}%
\bibitem [{SM()}]{SM}%
  \BibitemOpen
  \href@noop {} {\ }\bibinfo {note} {See Supplementary Material for
  model-parameter dependence of Ni density of states, hybridization intensity,
  and Ni 2p XPS spectra.}\BibitemShut {Stop}%
\bibitem [{\citenamefont {de~Groot}\ and\ \citenamefont
  {Kotani}(2014)}]{groot_kotani}%
  \BibitemOpen
  \bibfield  {author} {\bibinfo {author} {\bibfnamefont {F.}~\bibnamefont
  {de~Groot}}\ and\ \bibinfo {author} {\bibfnamefont {A.}~\bibnamefont
  {Kotani}},\ }\href {\doibase 10.1201/9781420008425} {\emph {\bibinfo {title}
  {Core Level Spectroscopy of Solids}}}\ (\bibinfo  {publisher} {CRC Press,
  Boca Raton, FL},\ \bibinfo {year} {2014})\BibitemShut {NoStop}%
\bibitem [{\citenamefont {Hirayama}\ \emph {et~al.}(2020)\citenamefont
  {Hirayama}, \citenamefont {Tadano}, \citenamefont {Nomura},\ and\
  \citenamefont {Arita}}]{Hirayama20}%
  \BibitemOpen
  \bibfield  {author} {\bibinfo {author} {\bibfnamefont {M.}~\bibnamefont
  {Hirayama}}, \bibinfo {author} {\bibfnamefont {T.}~\bibnamefont {Tadano}},
  \bibinfo {author} {\bibfnamefont {Y.}~\bibnamefont {Nomura}}, \ and\ \bibinfo
  {author} {\bibfnamefont {R.}~\bibnamefont {Arita}},\ }\href {\doibase
  10.1103/PhysRevB.101.075107} {\bibfield  {journal} {\bibinfo  {journal}
  {Phys. Rev. B}\ }\textbf {\bibinfo {volume} {101}},\ \bibinfo {pages}
  {075107} (\bibinfo {year} {2020})}\BibitemShut {NoStop}%
\bibitem [{\citenamefont {Ghijsen}\ \emph {et~al.}(1988)\citenamefont
  {Ghijsen}, \citenamefont {Tjeng}, \citenamefont {van Elp}, \citenamefont
  {Eskes}, \citenamefont {Westerink}, \citenamefont {Sawatzky},\ and\
  \citenamefont {Czyzyk}}]{Ghijsen88}%
  \BibitemOpen
  \bibfield  {author} {\bibinfo {author} {\bibfnamefont {J.}~\bibnamefont
  {Ghijsen}}, \bibinfo {author} {\bibfnamefont {L.~H.}\ \bibnamefont {Tjeng}},
  \bibinfo {author} {\bibfnamefont {J.}~\bibnamefont {van Elp}}, \bibinfo
  {author} {\bibfnamefont {H.}~\bibnamefont {Eskes}}, \bibinfo {author}
  {\bibfnamefont {J.}~\bibnamefont {Westerink}}, \bibinfo {author}
  {\bibfnamefont {G.~A.}\ \bibnamefont {Sawatzky}}, \ and\ \bibinfo {author}
  {\bibfnamefont {M.~T.}\ \bibnamefont {Czyzyk}},\ }\href {\doibase
  10.1103/PhysRevB.38.11322} {\bibfield  {journal} {\bibinfo  {journal} {Phys.
  Rev. B}\ }\textbf {\bibinfo {volume} {38}},\ \bibinfo {pages} {11322}
  (\bibinfo {year} {1988})}\BibitemShut {NoStop}%
\bibitem [{\citenamefont {Nomura}\ \emph {et~al.}(2020)\citenamefont {Nomura},
  \citenamefont {Nomoto}, \citenamefont {Hirayama},\ and\ \citenamefont
  {Arita}}]{Nomura20}%
  \BibitemOpen
  \bibfield  {author} {\bibinfo {author} {\bibfnamefont {Y.}~\bibnamefont
  {Nomura}}, \bibinfo {author} {\bibfnamefont {T.}~\bibnamefont {Nomoto}},
  \bibinfo {author} {\bibfnamefont {M.}~\bibnamefont {Hirayama}}, \ and\
  \bibinfo {author} {\bibfnamefont {R.}~\bibnamefont {Arita}},\ }\href
  {\doibase 10.1103/PhysRevResearch.2.043144} {\bibfield  {journal} {\bibinfo
  {journal} {Phys. Rev. Research}\ }\textbf {\bibinfo {volume} {2}},\ \bibinfo
  {pages} {043144} (\bibinfo {year} {2020})}\BibitemShut {NoStop}%
\bibitem [{\citenamefont {Nomura}\ \emph {et~al.}(2019)\citenamefont {Nomura},
  \citenamefont {Hirayama}, \citenamefont {Tadano}, \citenamefont {Yoshimoto},
  \citenamefont {Nakamura},\ and\ \citenamefont {Arita}}]{Nomura19}%
  \BibitemOpen
  \bibfield  {author} {\bibinfo {author} {\bibfnamefont {Y.}~\bibnamefont
  {Nomura}}, \bibinfo {author} {\bibfnamefont {M.}~\bibnamefont {Hirayama}},
  \bibinfo {author} {\bibfnamefont {T.}~\bibnamefont {Tadano}}, \bibinfo
  {author} {\bibfnamefont {Y.}~\bibnamefont {Yoshimoto}}, \bibinfo {author}
  {\bibfnamefont {K.}~\bibnamefont {Nakamura}}, \ and\ \bibinfo {author}
  {\bibfnamefont {R.}~\bibnamefont {Arita}},\ }\href {\doibase
  10.1103/PhysRevB.100.205138} {\bibfield  {journal} {\bibinfo  {journal}
  {Phys. Rev. B}\ }\textbf {\bibinfo {volume} {100}},\ \bibinfo {pages}
  {205138} (\bibinfo {year} {2019})}\BibitemShut {NoStop}%
\bibitem [{\citenamefont {Kune\ifmmode~\check{s}\else \v{s}\fi{}}\ \emph
  {et~al.}(2007{\natexlab{b}})\citenamefont {Kune\ifmmode~\check{s}\else
  \v{s}\fi{}}, \citenamefont {Anisimov}, \citenamefont {Lukoyanov},\ and\
  \citenamefont {Vollhardt}}]{kunes07a}%
  \BibitemOpen
  \bibfield  {author} {\bibinfo {author} {\bibfnamefont {J.}~\bibnamefont
  {Kune\ifmmode~\check{s}\else \v{s}\fi{}}}, \bibinfo {author} {\bibfnamefont
  {V.~I.}\ \bibnamefont {Anisimov}}, \bibinfo {author} {\bibfnamefont {A.~V.}\
  \bibnamefont {Lukoyanov}}, \ and\ \bibinfo {author} {\bibfnamefont
  {D.}~\bibnamefont {Vollhardt}},\ }\href {\doibase 10.1103/PhysRevB.75.165115}
  {\bibfield  {journal} {\bibinfo  {journal} {Phys. Rev. B}\ }\textbf {\bibinfo
  {volume} {75}},\ \bibinfo {pages} {165115} (\bibinfo {year}
  {2007}{\natexlab{b}})}\BibitemShut {NoStop}%
\bibitem [{\citenamefont {Lin}\ \emph {et~al.}(2021)\citenamefont {Lin},
  \citenamefont {Villar~Arribi}, \citenamefont {Fabbris}, \citenamefont
  {Botana}, \citenamefont {Meyers}, \citenamefont {Miao}, \citenamefont {Shen},
  \citenamefont {Mazzone}, \citenamefont {Feng}, \citenamefont
  {Chiuzb\ifmmode~\u{a}\else \u{a}\fi{}ian}, \citenamefont {Nag}, \citenamefont
  {Walters}, \citenamefont {Garc\'{\i}a-Fern\'andez}, \citenamefont {Zhou},
  \citenamefont {Pelliciari}, \citenamefont {Jarrige}, \citenamefont
  {Freeland}, \citenamefont {Zhang}, \citenamefont {Mitchell}, \citenamefont
  {Bisogni}, \citenamefont {Liu}, \citenamefont {Norman},\ and\ \citenamefont
  {Dean}}]{Lin2021}%
  \BibitemOpen
  \bibfield  {author} {\bibinfo {author} {\bibfnamefont {J.~Q.}\ \bibnamefont
  {Lin}}, \bibinfo {author} {\bibfnamefont {P.}~\bibnamefont {Villar~Arribi}},
  \bibinfo {author} {\bibfnamefont {G.}~\bibnamefont {Fabbris}}, \bibinfo
  {author} {\bibfnamefont {A.~S.}\ \bibnamefont {Botana}}, \bibinfo {author}
  {\bibfnamefont {D.}~\bibnamefont {Meyers}}, \bibinfo {author} {\bibfnamefont
  {H.}~\bibnamefont {Miao}}, \bibinfo {author} {\bibfnamefont {Y.}~\bibnamefont
  {Shen}}, \bibinfo {author} {\bibfnamefont {D.~G.}\ \bibnamefont {Mazzone}},
  \bibinfo {author} {\bibfnamefont {J.}~\bibnamefont {Feng}}, \bibinfo {author}
  {\bibfnamefont {S.~G.}\ \bibnamefont {Chiuzb\ifmmode~\u{a}\else
  \u{a}\fi{}ian}}, \bibinfo {author} {\bibfnamefont {A.}~\bibnamefont {Nag}},
  \bibinfo {author} {\bibfnamefont {A.~C.}\ \bibnamefont {Walters}}, \bibinfo
  {author} {\bibfnamefont {M.}~\bibnamefont {Garc\'{\i}a-Fern\'andez}},
  \bibinfo {author} {\bibfnamefont {K.-J.}\ \bibnamefont {Zhou}}, \bibinfo
  {author} {\bibfnamefont {J.}~\bibnamefont {Pelliciari}}, \bibinfo {author}
  {\bibfnamefont {I.}~\bibnamefont {Jarrige}}, \bibinfo {author} {\bibfnamefont
  {J.~W.}\ \bibnamefont {Freeland}}, \bibinfo {author} {\bibfnamefont
  {J.}~\bibnamefont {Zhang}}, \bibinfo {author} {\bibfnamefont {J.~F.}\
  \bibnamefont {Mitchell}}, \bibinfo {author} {\bibfnamefont {V.}~\bibnamefont
  {Bisogni}}, \bibinfo {author} {\bibfnamefont {X.}~\bibnamefont {Liu}},
  \bibinfo {author} {\bibfnamefont {M.~R.}\ \bibnamefont {Norman}}, \ and\
  \bibinfo {author} {\bibfnamefont {M.~P.~M.}\ \bibnamefont {Dean}},\ }\href
  {\doibase 10.1103/PhysRevLett.126.087001} {\bibfield  {journal} {\bibinfo
  {journal} {Phys. Rev. Lett.}\ }\textbf {\bibinfo {volume} {126}},\ \bibinfo
  {pages} {087001} (\bibinfo {year} {2021})}\BibitemShut {NoStop}%
\bibitem [{\citenamefont {{Zeng}}\ \emph {et~al.}(2021)\citenamefont {{Zeng}},
  \citenamefont {{Li}}, \citenamefont {{Chow}}, \citenamefont {{Cao}},
  \citenamefont {{Zhang}}, \citenamefont {{Tang}}, \citenamefont {{Yin}},
  \citenamefont {{Lim}}, \citenamefont {{Hu}}, \citenamefont {{Yang}},\ and\
  \citenamefont {{Ariando}}}]{Zeng21}%
  \BibitemOpen
  \bibfield  {author} {\bibinfo {author} {\bibfnamefont {S.~W.}\ \bibnamefont
  {{Zeng}}}, \bibinfo {author} {\bibfnamefont {C.~J.}\ \bibnamefont {{Li}}},
  \bibinfo {author} {\bibfnamefont {L.~E.}\ \bibnamefont {{Chow}}}, \bibinfo
  {author} {\bibfnamefont {Y.}~\bibnamefont {{Cao}}}, \bibinfo {author}
  {\bibfnamefont {Z.~T.}\ \bibnamefont {{Zhang}}}, \bibinfo {author}
  {\bibfnamefont {C.~S.}\ \bibnamefont {{Tang}}}, \bibinfo {author}
  {\bibfnamefont {X.~M.}\ \bibnamefont {{Yin}}}, \bibinfo {author}
  {\bibfnamefont {Z.~S.}\ \bibnamefont {{Lim}}}, \bibinfo {author}
  {\bibfnamefont {J.~X.}\ \bibnamefont {{Hu}}}, \bibinfo {author}
  {\bibfnamefont {P.}~\bibnamefont {{Yang}}}, \ and\ \bibinfo {author}
  {\bibfnamefont {A.}~\bibnamefont {{Ariando}}},\ }\href@noop {} {\  (\bibinfo
  {year} {2021})},\ \Eprint {http://arxiv.org/abs/2105.13492}
  {arXiv:2105.13492} \BibitemShut {NoStop}%
\bibitem [{\citenamefont {{Osada}}\ \emph {et~al.}(2021)\citenamefont
  {{Osada}}, \citenamefont {{Wang}}, \citenamefont {{Goodge}}, \citenamefont
  {{Harvey}}, \citenamefont {{Lee}}, \citenamefont {{Li}}, \citenamefont
  {{Kourkoutis}},\ and\ \citenamefont {{Hwang}}}]{Osada21}%
  \BibitemOpen
  \bibfield  {author} {\bibinfo {author} {\bibfnamefont {M.}~\bibnamefont
  {{Osada}}}, \bibinfo {author} {\bibfnamefont {B.~Y.}\ \bibnamefont {{Wang}}},
  \bibinfo {author} {\bibfnamefont {B.~H.}\ \bibnamefont {{Goodge}}}, \bibinfo
  {author} {\bibfnamefont {S.~P.}\ \bibnamefont {{Harvey}}}, \bibinfo {author}
  {\bibfnamefont {K.}~\bibnamefont {{Lee}}}, \bibinfo {author} {\bibfnamefont
  {D.}~\bibnamefont {{Li}}}, \bibinfo {author} {\bibfnamefont {L.~F.}\
  \bibnamefont {{Kourkoutis}}}, \ and\ \bibinfo {author} {\bibfnamefont
  {H.~Y.}\ \bibnamefont {{Hwang}}},\ }\href@noop {} {\  (\bibinfo {year}
  {2021})},\ \Eprint {http://arxiv.org/abs/2105.13494} {arXiv:2105.13494}
  \BibitemShut {NoStop}%
\end{thebibliography}%
\end{document}